%% file: preprint.tex
\documentstyle[aps,prb,floats]{revtex}

\newcommand{\nn}{\nonumber\\}

\begin{document}

\def\half {{\textstyle{1\over 2}}}
\def\threehalf {{\textstyle{3\over 2}}}
\def\Vdir {V^{\rm dir}}
\def\Vexc {V^{\rm exc}}
\def\Vd {V^{\rm d}}
\def\Vm {V^{\rm m}}
\def\DeltaR {\Delta{\bf R}}
\def\dxy {\rm d_{\rm x^2-y^2}}
\def\px {{\rm p_{\rm x}}}
\def\py {{\rm p_{\rm y}}}
\def\bfQ {{\bf Q}}
\def\bfk {{\bf k}}
\def\Udd {U_{\rm dd}}
\def\Upp {U_{\rm pp}}
\def\Upd {U_{\rm pd}}
\def\ibar {\overline\imath}
\def\navg {\langle n\rangle}
\def\gampp {\Gamma^{\rm pp}}
\def\hf {\hfill\break}
\def\cbar {\overline c}
\def\onemat {{\bf 1}}
\def\gamph {\Gamma^{\rm ph} }
\def\gamphbar { \smash{\overline\Gamma}{\vphantom\Gamma}^{\rm ph} }
\def\gph {G^{\rm ph} }
\def\gpp {G^{\rm pp} }
\def\chiph {\chi^{\rm ph} }
\def\chipp {\chi^{\rm pp} }
\def\bfR {{\bf R}}
\def\sqrthalf {{\textstyle{1\over \sqrt{2}}}}
\def\fourth {{\textstyle{1\over 4}}}
\def\halfsqrthalf {{\textstyle{1\over {2\sqrt 2}}}}
\def\hitc {high-$T_{\rm c}$}
\def\cuotwo {$\rm CuO_{2}$}
\def\tc {T_{\rm c}}
\def\tpd {t_{\rm pd}}
\def\tpp {t_{\rm pp}}
\def\epsd {\varepsilon_{\rm d}}
\def\epsp {\varepsilon_{\rm p}}
\def\eps {\varepsilon}
\def\udd {U_{\rm dd}}
\def\upp {U_{\rm pp}}
\def\upd {U_{\rm pd}}
\def\eV {\rm eV}
\def\kb {k_{\rm B}}
\def\etal {{\em et~al.\/}}
\def\ibid {{\em ibid.\/}}
\def\bsp {\!\!\!\!}

\draft

\title{FLUCTUATION EXCHANGE ANALYSIS OF SUPERCONDUCTIVITY\\
IN THE STANDARD THREE-BAND CuO$_2$ MODEL}
\author{G\"{o}khan Esirgen and N.~E. Bickers}
\address{Institute for Theoretical Physics\\University of California\\Santa Barbara, CA 93106-4030\\and\\Department of Physics and Astronomy\\University of Southern California\\Los Angeles, CA 90089-0484}

\maketitle

\begin{abstract}
\input{abs.tex}
\end{abstract}


\def\ratio {2\Delta(0)/k_BT_c}
\def\ratiok {2\Delta_{\bf k}(0)/k_BT_c}
\def\potent {V_{\rm sing}({\bf k},\thinspace {\bf k}') }
\def\chin {\bar{\chi}({\bf k}-{\bf k}') }
\def\chins {\bar{\chi}_s({\bf k}-{\bf k}') }
\def\chinc {\bar{\chi}_c({\bf k}-{\bf k}') }
\def\Ek {E_{\bf k} }
\def\Ekq {E_{\bf k+q} }
\def\uk {u_{\bf k} }
\def\ukq {u_{\bf k+q} }
\def\vk {v_{\bf k} }
\def\vkq {v_{\bf k+q} }

\section{Introduction}
\label{ch4int}
  \input{int.tex}

\section{Model and Notation}
\label{ch4mod}
  \input{mod.tex}

\section{Derivation of Vertex Functions}
\label{ch4der}
  \input{der.tex}

\section{Results and Discussion}
\label{ch4res}
  \input{res.tex}

\section{Summary}
\label{ch4sum}
  \input{sum.tex}

\acknowledgements
\input{ack.tex}

\input{ref.tex}
\end{document}

%% file: abs.tex
The fluctuation exchange, or FLEX, approximation for interacting electrons
is applied to study instabilities in the standard three-band model for
{\cuotwo} layers in the high-temperature superconductors.
Both intra-orbital and near-neigbor Coulomb interactions are retained.
The filling dependence of the $\dxy$ transition temperature
is studied in both the ``hole-doped'' and ``electron-doped'' regimes using
parameters derived from constrained-occupancy density-functional theory
for $\rm{La}_{2}\rm{CuO}_{4}$. The agreement with experiment on the overdoped
hole side of the phase diagram is remarkably good, i.e., transitions
emerge in the $40\:\rm{K}$ range with no free parameters. In addition the
importance of the ``orbital antiferromagnetic,'' or flux phase, charge
density channel is emphasized for an understanding of the underdoped regime.

%% file: int.tex
An experimental consensus has developed in recent years that the order parameter in
the high-temperature cuprate superconductors has $\dxy$ symmetry.\cite{tr1} Well before experiments
indicated this exotic symmetry a variety of theoretical approaches had suggested a tendency
toward $\dxy$ pairing in the Hubbard\cite{tr2,tr3,tr4} and $t-J$ models.\cite{tr5} Within weak-coupling 
approaches,
which treat the Coulomb interaction as a perturbation to one-electron band theory, 
exchange of antiferromagnetic spin fluctuations\cite{tr6} leads to pairing.

While the correctness of the spin fluctuation scenario remains controversial, it is of
interest to examine the pairing process within a more realistic setting than the one-band
Hubbard model. It is well-established that magnetism in the ``undoped'' cuprates can
be understood within the context of a three-band model\cite{tr7} (which projects to a $t-J$ model\cite{tr8}
in the strong-coupling limit). This CuO$_2$ model describes nearly filled Cu 3$\dxy$,
O 2p$_{x}$, and O 2p$_{y}$ orbitals, which form a two-dimensional square Bravais
lattice with a three-atom unit cell. The largest Coulomb integrals\cite{tr9,tr10,tr11} in 
the CuO$_2$ model
are the repulsion between holes on the same d orbital ($U_{\rm dd}\sim 10$ eV)
or p orbital ($U_{\rm pp}\sim 4$ eV), and the repulsion between holes on neighboring
d and p orbitals ($U_{\rm pd}\sim 1$ eV). 

A self-consistent and conserving calculation of one-particle properties in the
CuO$_2$ model based on exchange of magnetic and charge density fluctuations has been carried
out previously.\cite{tr12,tr13} In the present paper we extend this fluctuation exchange, or ``FLEX,''
calculation to an analysis of eigenvalues of the particle-particle and particle-hole
vertex functions and the
resulting transition temperatures. In particular, this analysis is
carried out using one- and two-particle matrix elements deduced from constrained-occupancy
density functional
theory,\cite{tr9} with no additional model projections or parameter fits. 
The results of this calculation
with no adjustable parameters are, if not compelling, at least suggestive.

While the FLEX approach is inherently approximate, the observed trends in eigenvalues and 
transition
temperatures for variations in hole density and Coulomb integrals can be expected
to be carried over in more exact treatments. In addition this calculation provides a
detailed example of the melding of many-body and band theory techniques now possible.

The paper is organized as follows: The model and calculational notation are summarized in
Section~\ref{ch4mod}. The particle-particle and particle-hole vertex functions within the 
FLEX approximation are
derived in a computationally tractable form in Section~\ref{ch4der}. After a brief digression on
sources of error, results for eigenvalues, transition temperatures, and
eigenfunctions are presented in
Section~\ref{ch4res}. The implications of the calculation are discussed, along with an
overall summary, in Section~\ref{ch4sum}.

%% file: mod.tex
\begin{figure}[hbtp]
\begin{picture}(80,65)
\includegraphics{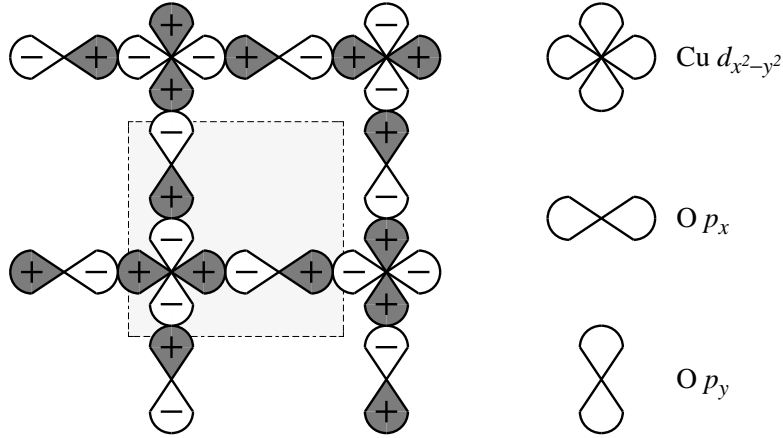}
\end{picture}
\caption{Unit cell and orbital phase conventions. The unit cell contains three
orbitals: the Cu 3d$_{\rm x^2-y^2}$, the O 2p$_{\rm x}$ on x-axis bonds, and
the O 2p$_{\rm y}$ on y-axis bonds. The orbital phases are chosen in a checkerboard
pattern. This assures that near-neighbor Cu--O and O--O hopping integrals have the
same sign in all unit cells and greatly simplifies the two-body eigenstate analysis.}
\label{tf1}
\end{figure}

In this section we define our notational conventions for the model to be studied. The three-orbital
model for superconducting cuprate layers may be written in terms of creation operators for
holes or electrons. As in Reference~\onlinecite{tr12}, which we hereafter denote ``EB,'' we 
adopt the hole representation; as an example, 
$c^{\dag}_{\rm d
\sigma}(\bfR)$ creates a 3d$_{\rm x^2-y^2}$ hole with spin $\sigma$ in unit cell $\bfR$. In addition
we choose a staggered orbital phase which helps simplify the analysis of two-particle eigenstates.
The unit cell and phase conventions are illustrated in Figure~\ref{tf1}.
The Hamiltonian
is conveniently broken up into one-particle and two-particle components
\begin{equation}
\label{te1}
\widehat H\ =\ \widehat H_0 + \widehat V\ .
\end{equation}
With our conventions the one-particle Hamiltonian takes the form
\begin{eqnarray}
\label{te2}
 \lefteqn{\widehat H_0-\mu N\ =}\nn&&(\epsd-\mu)\sum_{\bf R}\,n_{\rm d}
  ({\bf R})\ +\ (\epsd-\mu+\varepsilon)\sum_{\bf R}\,\Bigl[n_{\rm x}
  ({\bf R})+n_{\rm y}({\bf R})\Bigr]\nn&&
  \mbox{}-\tpd\sum_{\sigma,\,{\bf R}}\,
  \Bigl[\,c^{\dag}_{{\rm d}\sigma}({\bf R})c^{\vphantom{\dag}}_{{\rm x}
  \sigma}({\bf R})\,+\,
  c^{\dag}_{{\rm d}\sigma}({\bf R}+\widehat x)c^{\vphantom{\dag}}_{{\rm x}
  \sigma}({\bf R})\nn&&
  \mbox{}+\,c^{\dag}_{{\rm d}\sigma}({\bf R})c^{\vphantom{\dag}}_{{\rm y}
  \sigma}({\bf R})\,+\,
  c^{\dag}_{{\rm d}\sigma}({\bf R}+\widehat y)c^{\vphantom{\dag}}_{{\rm y}
  \sigma}({\bf R})\,+\,H.C.\,\Bigr]\nn&&
  \mbox{}-\tpp\sum_{\sigma,\,{\bf R}}\,
  \Bigl[\,c^{\dag}_{{\rm y}\sigma}({\bf R})c^{\vphantom{\dag}}_{{\rm x}
  \sigma}({\bf R})\,+\,
  c^{\dag}_{{\rm y}\sigma}({\bf R}+\widehat x)c^{\vphantom{\dag}}_{{\rm x}
  \sigma}({\bf R})\nn&&
  \mbox{}+\,c^{\dag}_{{\rm y}\sigma}({\bf R}-\widehat y)
  c^{\vphantom{\dag}}_{{\rm x}\sigma}({\bf R})\,+\,
  c^{\dag}_{{\rm y}\sigma}({\bf R}+\widehat x-\widehat y)
  c^{\vphantom{\dag}}_{{\rm x}\sigma}({\bf R})\,+\,H.C.\,\Bigr]\ .
\end{eqnarray}
The number operators $n_{\rm d}$, $n_{\rm x}$ and $n_{\rm y}$ are defined in the
usual way, e.g.,
\begin{equation}
\label{te3}
n_{\rm d}(\bfR)\ =\ \sum_{\sigma}\,c_{\rm d\sigma}^{\dag}(\bfR)c^{\vphantom{\dag}}
_{\rm d\sigma}(\bfR)\ .
\end{equation}
The physical values of the short-range hopping matrix elements $\tpd$ and $\tpp$ are both 
positive.\cite{tr9,tr10}
The d-hole creation energy $\epsd$ may be set to zero without loss, and the
p--d energy level difference $\varepsilon$ is positive. 

In the two-body Hamiltonian $\widehat V$ we retain the three largest Coulomb integrals
from constrained-occupancy density functional studies:\cite{tr9,tr10} The on-site 
Cu repulsion $\Udd$, the on-site O
repulsion $\Upp$, and the near-neighbor Cu--O repulsion $\Upd$. The last interaction
complicates the analysis since it has both intra-cell and inter-cell components. The
Coulomb interactions may be written in a spin-diagonalized form which allows a decoupling
of $S=0$ (density) and $S=1$ (magnetic) excitations. This procedure is treated at length in
EB.

\begin{figure}[hbtp]
\begin{picture}(80,25)
\includegraphics{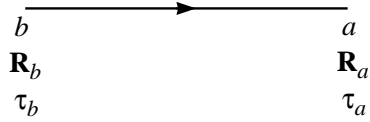}
\end{picture}
\caption{Diagrammatic representation of the one-particle propagator
$G_{ab}(\DeltaR_{ab},\,\Delta\tau_{ab})$.}
\label{tf2}
\end{figure}

The one-particle propagators for the Hamiltonian described above take the form
\begin{equation}
\label{te4}
G_{ab}(\bfR_a,\,\tau_a;\,\bfR_b,\,\tau_b)\ =\ G_{ab}(\Delta\bfR_{ab},\,\Delta\tau_{ab})
  \ \equiv\ -\langle T_{\tau}c^{\vphantom{\dag}}_a(\bfR_a,\,\tau_a)c^{\dag}_b(\bfR_b,\,\tau_b)
  \rangle\ ,
\end{equation}
where $(a,\,\bfR_a,\,\tau_a)$ and $(b,\,\bfR_b,\,\tau_b)$ are the orbital, unit-cell, and 
imaginary-time labels for particles in the final and initial states (see Figure~\ref{tf2}).
Reference~\onlinecite{tr12}
describes the general procedure for calculating such propagators and provides detailed results
for the CuO$_2$ model described above. For the remainder of this paper we make use of
one-particle properties obtained in this previous study.

%% file: der.tex
\begin{figure}[hbtp]
\begin{picture}(80,65)
\includegraphics{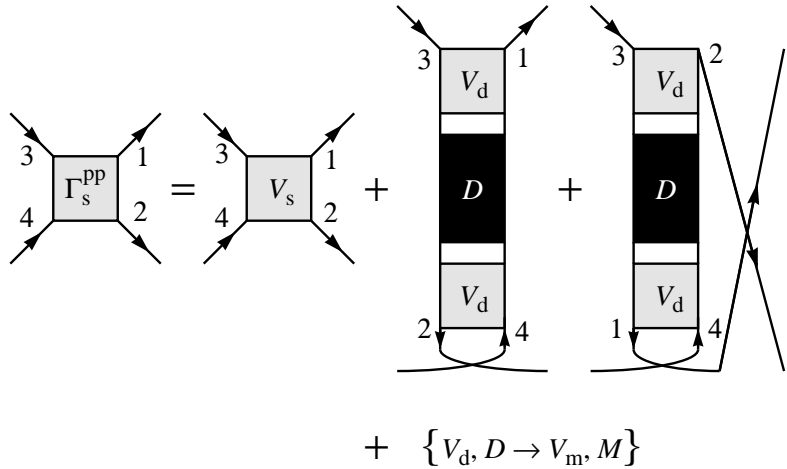}
\end{picture}
\caption{Irreducible singlet vertex function $\Gamma^{\rm pp}_{\rm s}$ 
within the FLEX approximation. Outgoing
states are represented on the right of the diagrams, incoming states on the left. (The
coefficients $1/2$ and $-3/2$ are omitted for clarity; see Equation~(\protect\ref{te5}).) 
$V_{\rm s}$ is the unrenormalized Coulomb matrix element in the singlet channel. The vertical
ladders represent the exchange of density fluctuations.}
\label{tf3}
\end{figure}

The calculation of eigenvalues of the particle-particle kernel in the CuO$_2$
model is conceptually straightforward, but notationally involved. It is
assumed that self-consistent one-particle propagators $G$ have been
obtained using the technique described in EB. Functional
differentiation of the off-diagonal self-energy in the presence of
an external pairing field yields the irreducible particle-particle vertex
$\gampp$. Using the notation developed in Reference~\onlinecite{tr14} the singlet and
triplet parts of the vertex are as follows (see Figure~\ref{tf3}):
\begin{eqnarray}
\label{te5}
\lefteqn{ \gampp_{\rm s}(12;\,34)
  \ = \ V_{\rm s}(12;\,34)} \nn & & \mbox{}\ + \ \half\Phi_{\rm d}(24;\,31)\ -\ \threehalf
  \Phi_{\rm m}(24;\,31)\ +\ \half\Phi_{\rm d}(14;\,32)
  \ -\ \threehalf\Phi_{\rm m}(14;\,32)\ , \\
\label{te6}
\lefteqn{ \gampp_{\rm t}(12;\,34)
\  =\ V_{\rm t}(12;\,34)} \nn & & \mbox{}\ + \ \half\Phi_{\rm d}(24;\,31)\ +\ \half
  \Phi_{\rm m}(24;\,31)\ -\ \half\Phi_{\rm d}(14;\,32)
  \ -\ \half\Phi_{\rm m}(14;\,32)\ . 
\end{eqnarray}
The numerical indices represent the space and time degrees of freedom of
each particle, i.e.,
\begin{equation}
\label{te7}
 1\ \equiv\ (\,m_1,\,{\bf R}_1,\,\tau_1\,)\ , 
\end{equation}
with $m_1$ the orbital, ${\bf R}_1$ the unit-cell displacement, and
$\tau_1$ the imaginary time coordinate for particle 1.

The matrix functions $\Phi_{\rm d}$ and $\Phi_{\rm m}$ represent particle-hole
ladders in the density and magnetic channels:
\begin{equation}
\label{te8}
 \Phi_{\rm r}(12;\,34)\ =\ \Bigl[V_{\rm r}\gph(\onemat- V_{\rm r}\gph)^{-1}
  V_{\rm r}\Bigr](12;\,34) 
\end{equation}
for $\rm r=d$ and $\rm m$. The matrices $V_{\rm r}$ are the spin-diagonalized
Coulomb interactions in each channel, and the matrix $\gph$ is the
uncorrelated particle-hole propagator:
\begin{eqnarray}
\label{te9}
 \gph(12;\,34)&=&\beta\thinspace
  \langle T_{\tau}c(1)c^{\dag}(3)\rangle\thinspace
  \langle T_{\tau}c(4)c^{\dag}(2)\rangle \nn
                        &=&\beta\, G(13)G(42)\ , 
\end{eqnarray}
with $\beta$ the inverse temperature. 

As usual,\cite{tr12} matrix multiplication is defined by
\begin{equation}
\label{te10}
 (AB)\,(12;\,34)\ =\ A(12;\,56)\,B(56;\,34)\ , 
\end{equation}
with an implied sum on repeated indices. The singlet and triplet kernels
are obtained by multiplying the vertex functions by the uncorrelated 
particle-particle propagator
\begin{equation}
\label{te11}
\gpp(12;\,34)\ =\ -\half\beta\, G(13)G(24) \ . 
\end{equation}
Note the presence of $\half$ in this definition of the propagator,
which is consistent with our normalization of the vertex functions below.

\begin{figure}[hbtp]
\begin{picture}(80,65)
\includegraphics{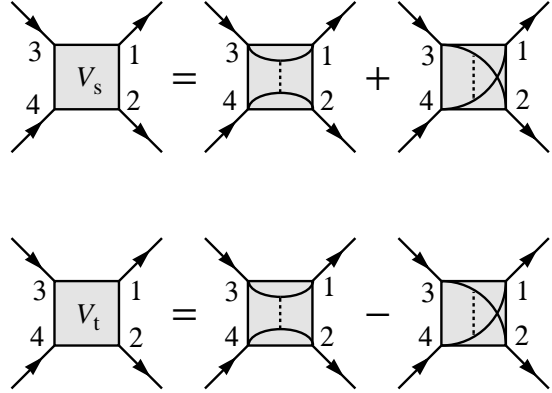}
\end{picture}
\caption{Representation of the unrenormalized singlet and triplet Coulomb
matrix elements $V_{\rm s}$ and $V_{\rm t}$.}
\label{tf4}
\end{figure}

\begin{figure}[hbtp]
\begin{picture}(80,35)
\includegraphics{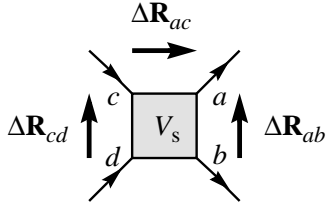}
\end{picture}
\caption{Definition of unit-cell displacements in the representation of
the singlet Coulomb matrix element.}
\label{tf5}
\end{figure}

Expressions for the density and magnetic Coulomb matrix elements
$V_{\rm d}$ and $V_{\rm m}$ have been given previously in EB.
Explicit expressions for $V_{\rm s}$ and $V_{\rm t}$ follow from the
diagrams in Figure~\ref{tf4}.
As in our previous work, it is convenient to adopt a
notation which emphasizes the dependence of the matrix elements on only three
unit-cell displacements between the two initial-state and two final-state
particles (see Figure~\ref{tf5}).
Thus, $V_{\rm s}(\Delta{\bf R}_{ac};\,ab,\,\Delta{\bf R}
  _{ab};\,cd,$ $\Delta{\bf R}_{cd})$ is the singlet Coulomb matrix element
for a final-state particle pair in orbitals $a$ and $b$ with
relative unit-cell displacement 
\begin{equation}
\label{te12}
\Delta{\bf R}_{ab}\ =\ {\bf R}_a - {\bf R}_b \ , 
\end{equation}
and an initial-state particle pair in orbitals $c$ and $d$ with
relative unit-cell displacement $\Delta{\bf R}_{cd}$.
The displacement between the initial and final state particles is
given by $\Delta{\bf R}_{ac}$.
There are only 11 two-particle states $\,(ab,\,\Delta{\bf R}_{ab}\,)$
which have non-zero singlet and triplet Coulomb matrix elements
in the CuO$_2$ model considered here. These states are listed in Table~\ref{tt1}
for the unit-cell depicted in Figure~\ref{tf1}. (An identically labeled 11-state
basis for non-zero density and magnetic Coulomb matrix elements was
defined in EB.) 
\begin{table}
\begin{center}
\begin{tabular}{|c|c|c|c|} \hline
Index & $a$ & $b$ & $\DeltaR_{ab}$ \\ \hline
1 & d & d & 0 \\
2 & $\px$ & $\px$ & 0 \\
3 & $\py$ & $\py$ & 0 \\
4 & d & $\px$ & 0 \\
5 & d & $\px$ & $+\widehat x$ \\
6 & $\px$ & d & 0 \\
7 & $\px$ & d & $-\widehat x$ \\
8 & d & $\py$ & 0 \\
9 & d & $\py$ & $+\widehat y$ \\
10 & $\py$ & d & 0 \\
11 & $\py$ & d & $-\widehat y$ \\ \hline
\end{tabular}
\end{center}
\caption{Indexing scheme for the minimum-range
particle-particle basis set in the 
CuO$_2$ model. The particle orbitals are $a$ and $b$, with corresponding 
unit-cell displacement $\Delta{\bf R}_{ab}\,\equiv\,{\bf R}_a-{\bf R}_b$.
Note that kernel eigenstates must satisfy the symmetry requirements of the
Pauli Principle, but the basis states need not.}
\label{tt1}
\end{table}

As in EB, the initial/final state displacement $\Delta{\bf R}_{ac}$ is conveniently 
eliminated in favor of a center-of-mass momentum $\bf Q$ by writing
\begin{equation}
\label{te13}
V_{\rm s}(\bfQ;\,ab,\,\Delta{\bf R}_{ab};\,cd,\,\Delta{\bf R}_{cd})\ \equiv
  \ \sum_{\Delta{\bf R}_{ac}}\,e^{-i\bfQ\cdot\Delta{\bf R}_{ac}}\,
  V_{\rm s}(\Delta{\bf R}_{ac};\,ab,\,\Delta{\bf R}_{ab};\,cd,\,\Delta{\bf R}_{cd})
  \ . 
\end{equation}
The indices in Table~\ref{tt1} can then be used to write $V_{\rm s}$ and
$V_{\rm t}$ compactly as $\bfQ$-dependent $11\times 11$ matrices. For
example,
\begin{eqnarray}
\label{te14}
 V_{\rm s}^{33}(\bfQ) &=& 2\Upp \nn
   V_{\rm s}^{64}(\bfQ) &=& -V_{\rm t}^{64}(\bfQ)\ =\ \Upd \nn
   V_{\rm s}^{75}(\bfQ) &=& -V_{\rm t}^{75}(\bfQ)\ =\ e^{iQ_x}\Upd\ .
\end{eqnarray}

\begin{figure}[hbtp]
\begin{picture}(160,160)
\includegraphics{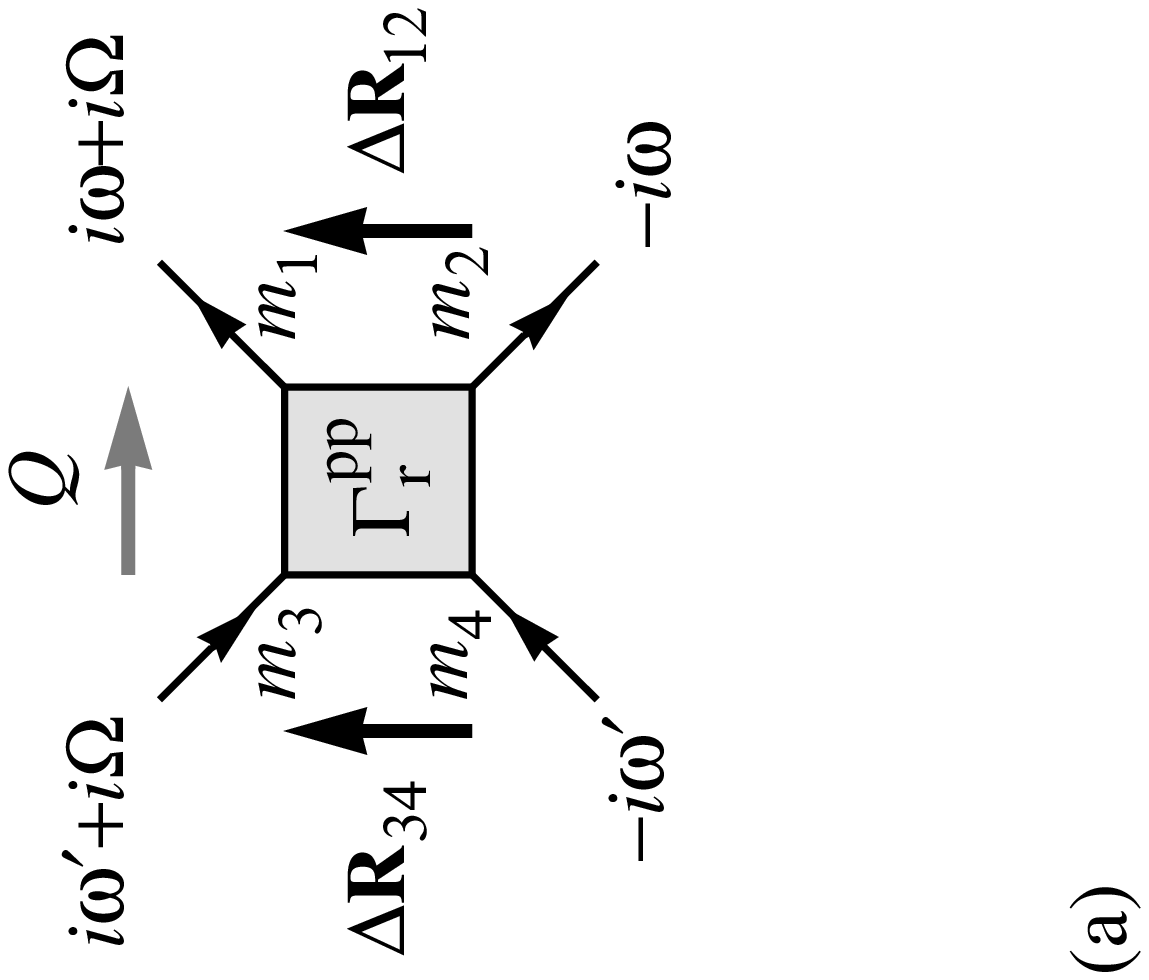}
\includegraphics{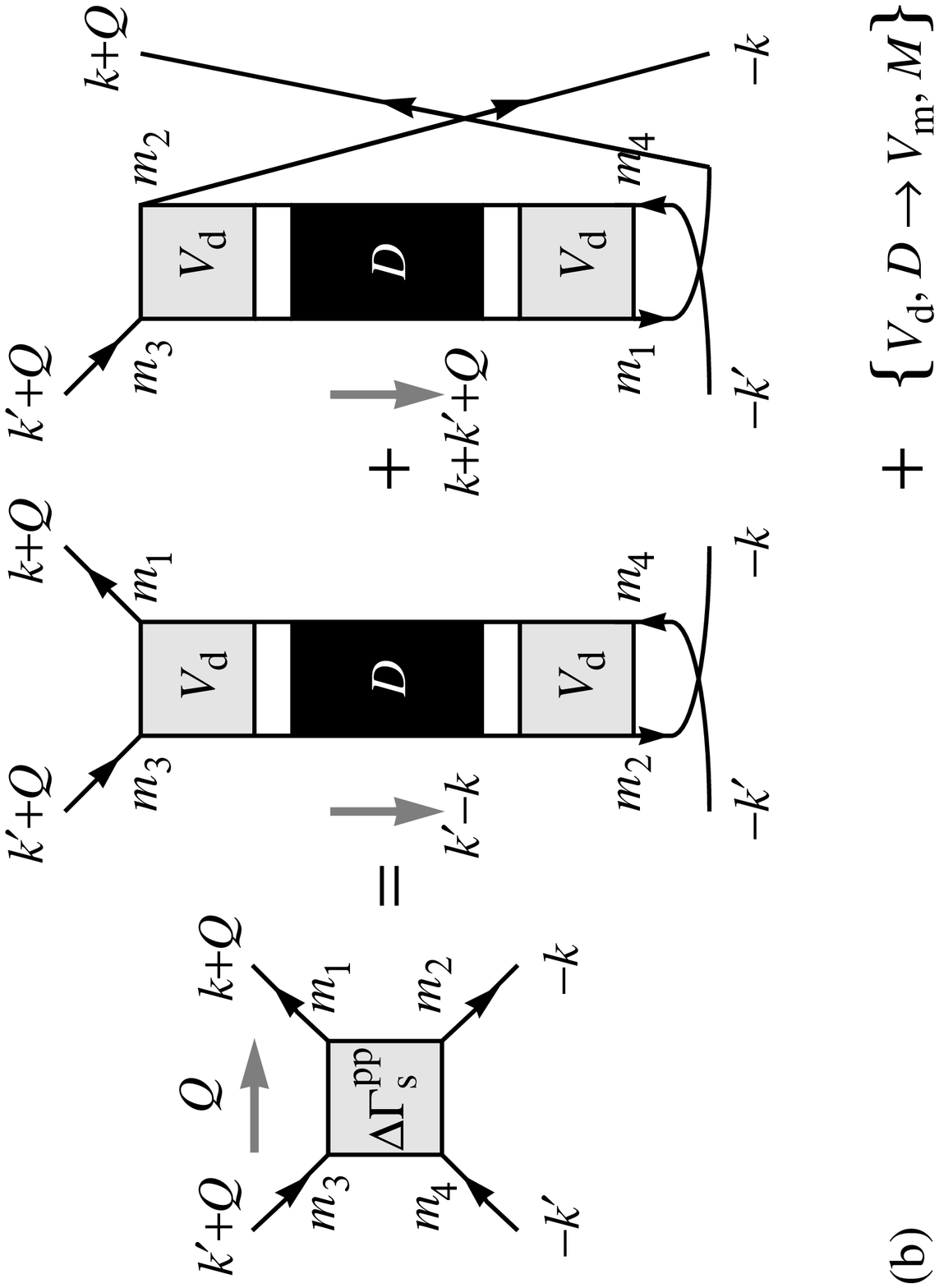}
\includegraphics{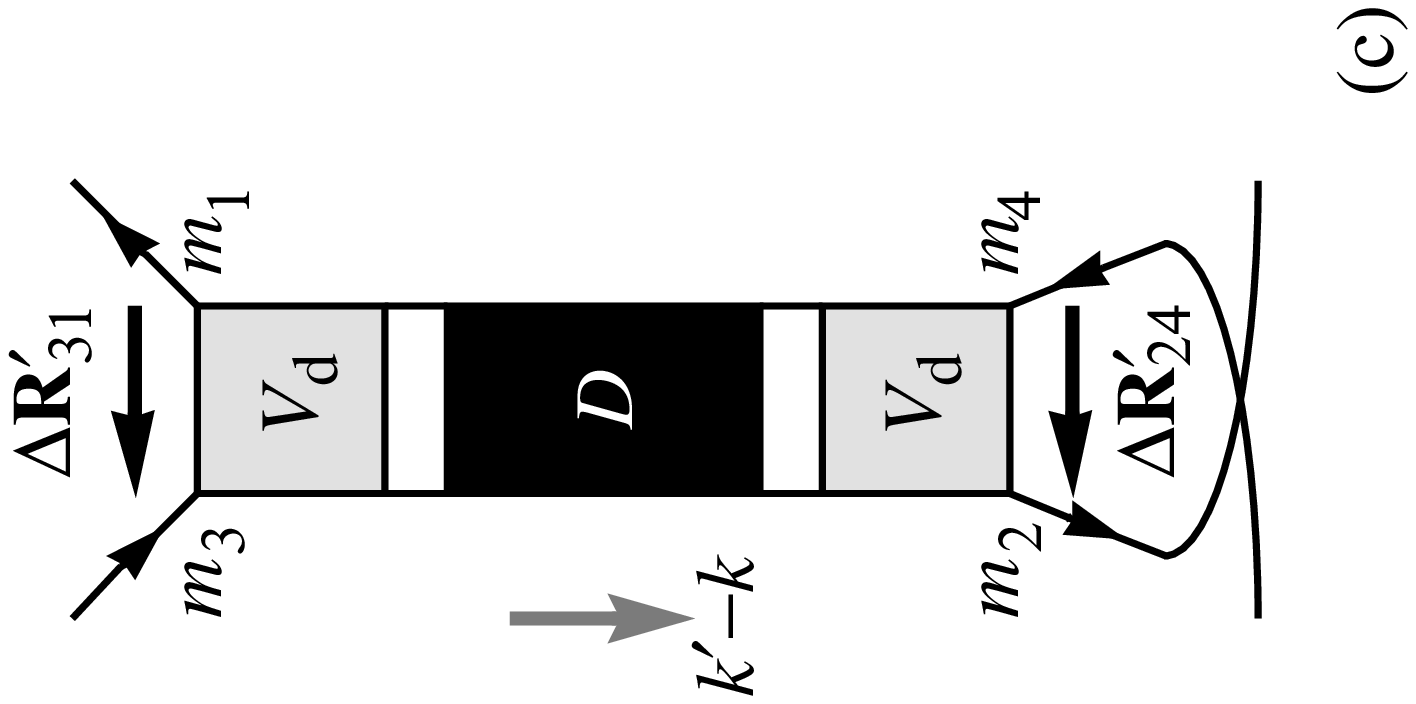}
\end{picture}
\caption{Calculation of the irreducible particle-particle vertex functions
$\Gamma^{\rm pp}_{\rm r}$, $\rm r=s$ and $\rm t$. (a)~Diagrammatic representation of
the irreducible vertex in the computationally optimal basis set. Note that the total
center-of-mass momentum-frequency $Q=(\bfQ,\,i\Omega)$ is conserved. (b)~Fourier-transformed
singlet vertex function $\Delta\Gamma^{\rm pp}_{\rm s}(Q;\,m_1m_2,\,k;\, m_3m_4,\,k')$. 
See also Equation~(\protect\ref{te18}).
(c)~Representation of the first particle-hole ladder in (b) in the relative displacement
basis. See also Equation~(\protect\ref{te19}).}
\label{tf6}
\end{figure}

Though the basic Coulomb interactions $V_{\rm s}$ and $V_{\rm t}$ are
short-ranged, the fluctuation-induced contributions to the particle-particle vertex
functions $\gampp_{\rm s}$ and $\gampp_{\rm t}$ are not. Nevertheless, 
it is possible to calculate accurate pairing eigenvalues
using vertex functions truncated in the relative displacement of
the particle pair. For this reason it is convenient to arrive at a
particle-particle vertex labeled using (i)~total momentum-frequency
$Q\,\equiv\,(\bfQ,\,i\Omega)$; (ii)~pair orbital indices $(ab)$
($3\times 3=9$ possible combinations for the three-orbital model); 
(iii)~unit-cell displacement $\Delta{\bf R}_{ab}$ of the pair elements;
and (iv)~relative frequency $i\omega$.\cite{tr15} (There is no additional benefit in
introducing a relative time coordinate, since the fluctuations
induce long-range couplings in imaginary time.) Previous notation
for the time-independent Coulomb matrix elements may be generalized
in a natural way. The desired singlet and triplet vertex functions (see Figure~\ref{tf6}(a)) 
take the form
\begin{equation}
\label{te15}
\gampp_{\rm r}(Q;\,m_1m_2,\,\Delta{\bf R}_{12},
  \,i\omega;\,
  m_3m_4,\,\Delta{\bf R}_{34},\,i\omega')\ . 
\end{equation}

In order to calculate the crossed-channel particle-hole ladders 
$\Phi_{\rm d}$ and
$\Phi_{\rm m}$, it is essential to use a different basis set obtained by a 
series of Fourier transforms. An initial Fourier transform on the
relative displacement coordinates in Equation~(\ref{te5}) yields
\begin{eqnarray}
\label{te16}
 \lefteqn{\gampp_{\rm s}(Q;\,m_1m_2,\,\Delta{\bf R}_{12},
  \,i\omega;\,
  m_3m_4,\,\Delta{\bf R}_{34},\,i\omega')\ =}\nn
&&V_{\rm s}(\bfQ;\,m_1m_2,\,\Delta{\bf R}_{12};
  \,m_3m_4,\,\Delta{\bf R}_{34})\nn
 &&\mbox{}\ +\ {1\over N^2}\sum_{\bfk\bfk'}\,e^{i\bfk\cdot\Delta{\bf R}_{12}}\,
  \Delta\gampp_{\rm s}(Q;\,m_1m_2,\,k;\,
  m_3m_4,\,k')\,e^{-i\bfk'\cdot\Delta{\bf R}_{34}}\ ,
\end{eqnarray}
with
\begin{eqnarray}
\label{te17}
 k&=&(\bfk,\,i\omega)\nn
           k'&=&(\bfk',\,i\omega')\ . 
\end{eqnarray}
The fluctuation-induced contribution $\Delta\gampp_{\rm s}$ takes the form
\begin{eqnarray}
\label{te18}
\lefteqn{ \Delta\gampp_{\rm s}(Q;\,m_1m_2,\,k;\,
  m_3m_4,\,k')\ =}\nn
   &&\Bigl[\half\Phi_{\rm d}-\threehalf\Phi_{\rm m}\Bigr]\,(k'-k;\,m_2m_4,\,-k';\,m_3m_1,
  \,k+Q)\nn
  &&\mbox{}\ +\ \Bigl[\half\Phi_{\rm d}-\threehalf\Phi_{\rm m}\Bigr]\,(k+k'+Q;\,m_1m_4,\,-k';\,m_3m_2,\,
  -k)\ .
\end{eqnarray}  
The ladders are represented diagrammatically in Figure~\ref{tf6}(b).

The first particle-hole ladder in Equation~(\ref{te18}) may be 
translated back to the relative displacement basis (Figure~\ref{tf6}(c)):
\begin{eqnarray}
\label{te19}
 \lefteqn{\Phi_{\rm d}(k'-k;
  \,m_2m_4,\,-k';\,m_3m_1,\,k+Q)\ =}\nn
&&\sum_{\Delta{\bf R}'_{24},\,\Delta{\bf R}'_{31}}\,
  e^{i\bfk'\cdot\Delta{\bf R}'_{24}}\,\Phi_{\rm d}(k'-k
  ;\,m_2m_4,\,\Delta{\bf R}'_{24};\,m_3m_1,\,\Delta{\bf R}'_{31})
  \,e^{i(\bfk+\bfQ)\cdot\Delta{\bf R}'_{31}}\ .\nn
\end{eqnarray}   
Note that primes are included on the displacements here to emphasize that
they are dummy summation variables, at this stage unrelated to $\Delta{\bf R}_{12}$ and 
$\Delta{\bf R}_{34}$ in Equation~(\ref{te16}).
Similar expressions hold for the other ladder sum terms contributing
to $\Delta\gampp_{\rm s}$.
Note that $\Phi_{\rm d}$ and $\Phi_{\rm m}$ are independent of the 
relative frequency variables
(due to the instantaneous character of $V_{\rm d}$ and $V_{\rm m}$) and have been
dropped from the notation without loss. 

The expressions in Equations~(\ref{te16}) and (\ref{te19}) each involve double Fourier transforms
and are impractical to calculate numerically. A much simpler form for $\Delta\Gamma$
may be derived by changing momentum variables and interchanging the order of sums. For example,
for the first particle-hole ladder contribution the appropriate change of variables is
\begin{eqnarray}
\label{te20}
 \bfk'-\bfk &\rightarrow& \bfQ' \nn
            \bfk' &\rightarrow& \bfk'\ . 
\end{eqnarray}
The sum on $\bfk'$ may then be carried out explicitly, yielding a delta function, which collapses the
sum on $\Delta{\bf R}'_{24}$.

After additional relabeling of summation variables,
the complete result for $\gampp_{\rm s}$ which results from this procedure is
\begin{eqnarray}
\label{te21}
\lefteqn{ \gampp_{\rm s}(Q;\,m_1m_2,\,\Delta{\bf R}_{12},\,i\omega;\,
  m_3m_4,\,\Delta{\bf R}_{34},\,i\omega')\ =}\nn
  &&V_{\rm s}(\bfQ;\,m_1m_2,\,\Delta{\bf R}_{12};
  \,m_3m_4,\,\Delta{\bf R}_{34})\nn  
   &&\mbox{}\ +\ \sum_{\Delta{\bf R}_{13}}\,e^{-i\bfQ\cdot\Delta{\bf R}_{13}}\,\Delta\gampp_{\rm s}
  (\Delta{\bf R}_{13},\,i\Omega;\,m_1m_2,\,\Delta{\bf R}_{12},\,i\omega;\,m_3m_4,\,\Delta{\bf R}_{34},\,
  i\omega')\ ,\nn
\end{eqnarray}
with
\begin{eqnarray}
\label{te22}
\lefteqn{ \Delta\gampp_{\rm s}
  (\Delta{\bf R}_{13},\,i\Omega;\,m_1m_2,\,\Delta{\bf R}_{12},\,i\omega;\,m_3m_4,\,\Delta{\bf R}_{34},\,
  i\omega')\ =} \nn
  &&{1\over N}\sum_{\bfQ'}\,e^{i\bfQ'\cdot\Delta{\bf R}_{23}}\,
   \Bigl[\half\Phi_{\rm d}-\threehalf\Phi_{\rm m}\Bigr](\bfQ',\,i(\omega'-\omega);\,m_2m_4,\, 
\Delta{\bf R}_{24};\,m_3m_1,\,\Delta{\bf R}_{31})\ + \nn
  &&{1\over N}\sum_{\bfQ'}\,e^{i\bfQ'\cdot\Delta{\bf R}_{13}}\,
   \Bigl[\half\Phi_{\rm d}-\threehalf\Phi_{\rm m}\Bigr](\bfQ',\,i(\omega+\omega'+\Omega);\,m_1m_4,\,
  \Delta{\bf R}_{14};\,m_3m_2,\,\Delta{\bf R}_{32})\ , \nn
\end{eqnarray}
where all relative displacements are expressed in terms of $\Delta{\bf R}_{13}$, $\Delta{\bf R}_{12}$,
and $\Delta{\bf R}_{34}$:
\begin{eqnarray}
\label{te23}
 \Delta{\bf R}_{23}&=&-\Delta{\bf R}_{32}\ =\ \Delta{\bf R}_{13}-\Delta{\bf R}_{12}
  \nn
  \Delta{\bf R}_{24}&=&\Delta{\bf R}_{13}-\Delta{\bf R}_{12}+\Delta{\bf R}_{34} \nn
  \Delta{\bf R}_{31}&=&-\Delta{\bf R}_{13} \nn
  \Delta{\bf R}_{14}&=&\Delta{\bf R}_{13}+\Delta{\bf R}_{34}\ . 
\end{eqnarray}
A similar expression for $\gampp_{\rm t}$ may be obtained immediately using the correspondence
in Equation~(\ref{te6}).

The ladder summations $\Phi_{\rm d}$ and $\Phi_{\rm m}$ may be calculated as matrix products in
the space with compound indices $(ab,\,\Delta{\bf R}_{ab})$: 
\begin{equation}
\label{te24}
\Phi_{\rm r}\ =\ -V_{\rm r}\overline\chi(\,1+V_{\rm r}\overline\chi\,)^{-1}V_{\rm r}\ ,
\end{equation}
for r=d, m, where the uncorrelated fluctuation propagator $\overline\chi$ is defined by
\begin{equation}
\label{te25}
\overline\chi(Q;\,ab,\,\Delta{\bf R}_{ab};\,
  cd,\,\Delta{\bf R}_{cd})\ =\ -{T\over N}\sum_{k}
  e^{i\bfk\cdot(\Delta{\bf R}_{ab}-\Delta{\bf R}_{cd})}\,
  G_{ac}(k+Q)G_{db}(k)\ . 
\end{equation}
For the CuO$_2$ model the required matrix inverse is only $11\times 11$.
Note, however, that a separate inverse must be calculated
for each value of the particle-hole ladder's center-of-mass 
momentum-frequency.

\begin{figure}[hbtp]
\begin{picture}(80,35)
\includegraphics{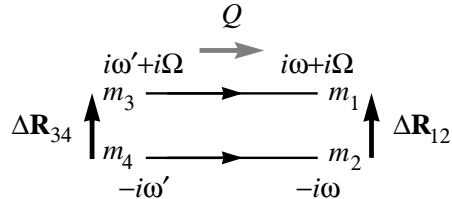}
\end{picture}
\caption{Diagrammatic representation of the uncorrelated particle-particle
propagator $G^{\rm pp}(Q;\,m_1m_2,\,\DeltaR_{12},\,i\omega;\,m_3m_4,\,\DeltaR_{34},\,i
\omega')$. Note the propagator is diagonal in the relative frequency, i.e., it vanishes
for $\omega\neq\omega'$.}
\label{tf7}
\end{figure}

The uncorrelated particle-particle propagator may also be
expressed in the basis adopted above (Figure~\ref{tf7}):
\begin{eqnarray}
\label{te26}
 \lefteqn{\gpp(Q;\,m_1m_2,\,\Delta{\bf R}_{12},\,i\omega;\,m_3m_4,\,
  \Delta{\bf R}_{34},\,i\omega')\ =}\nn
  &&-\half\delta_{\omega\omega'}
  {T\over N}\sum_{\bfk}\,e^{i\bfk\cdot(\Delta{\bf R}_{12}-\Delta
  {\bf R}_{34})}\,G_{m_1m_3}(k+Q)G_{m_2m_4}(-k)\ .
\end{eqnarray}
The particle-particle eigenvalue problem then takes the form
\begin{equation}
\label{te27}
\gampp_{\rm r}(Q)\gpp(Q)\phi(Q)\ =\ \lambda(Q)\phi(Q)\ , 
\end{equation}
for $\rm r=s$ and t. Note that with the conventions adopted here a positive eigenvalue
indicates attraction. (Although the kernel is non-hermitian,
it is possible to show for $\Omega=0$ that the particle-particle eigenvalues are
real-valued or occur in complex conjugate pairs.) The matrices $\gampp_{\rm r}$ and $\gpp$
operate in a far larger compound-index space than that defined previously for the Coulomb
interactions $V_{\rm r}$. The index now consists of the orbital-pair label 
$(m_1m_2)$, which takes on nine values in the CuO$_2$ problem; the subset of unit-cell displacements
$\Delta{\bf R}_{12}$ retained; and the set of values of the relative frequency $\omega$
within a pre-defined cutoff interval. 

Note that since the kernel matrix is
non-hermitian, its sets of left and right eigenvectors are not simply related (even though the
left and right eigenvalue spectra are identical.) In the following section we emphasize the
real-space and frequency dependence of the right eigenvectors, i.e., those determined 
by Equation~(\ref{te27}). This is natural since the right eigenvector at $T_c$ evolves smoothly into
the off-diagonal self-energy below $T_c$. (The right eigenvalue equation may be
re-derived by linearizing a self-consistent field problem in the off-diagonal
self-energy.) The corresponding left eigenvector has no such simple physical interpretation.

A number of powerful approaches have been developed
in recent years to compute a few selected
eigenvalues of a general non-hermitian matrix in 
cases such as this for which a full diagonalization is impractical.
All such approaches are derived from the much more standard algorithms
available for the real-symmetric and complex-hermitian eigenvalue problems.
We have made use of a so-called Lanczos-Arnoldi algorithm developed
in the Department of Computational and Applied Mathematics at Rice University.\cite{tr16} Using
this algorithm we have studied kernels with row dimensions of order 10,000. 

To supplement our study of particle-particle eigenvalues we have also
calculated a set of kernel eigenvalues for the particle-hole channels. These
channels describe scattering of $S=0$ (charge density) and $S=1$ (magnetic) excitations.
Several points are important to note in this regard. First of all, the FLEX calculation (and any
Baym-Kadanoff approach\cite{tr17}) lacks self-consistency at the two-particle level. 
For this reason the
density and magnetic propagators which enter the one-particle self-energy are not the same
as those obtained by functional differentiation of the self-energy with respect to an
external field. The difference may be described in terms of ``vertex corrections'' to the bare
density and magnetic matrix elements $V_{\rm d}$ and $V_{\rm m}$. Within FLEX the simplest
vertex corrections $\Delta\Gamma_{\rm d}$ and $\Delta\Gamma_{\rm m}$ have a form closely related
to the singlet and triplet interactions $\Delta\Gamma_{\rm s}$ and $\Delta\Gamma_{\rm t}$, i.e.,
they represent the exchange of {\sl single} crossed-channel density and magnetic fluctuations.
More complicated vertex corrections take the Aslamazov-Larkin (AL) form,\cite{tr18,tr19}
i.e., they describe the
emission and re-absorption of {\sl pairs} of fluctuations. For reasons described 
previously\cite{tr19}
we omit the AL corrections to $\Delta\Gamma_{\rm d}$ and $\Delta\Gamma_{\rm m}$ in the analysis
which follows.

\begin{figure}[hbtp]
\begin{picture}(80,55)
\includegraphics{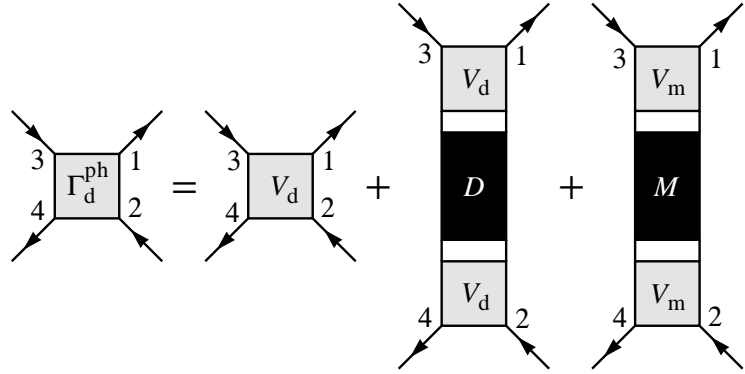}
\end{picture}
\caption{Irreducible density vertex function $\Gamma^{\rm ph}_{\rm d}$ 
within the FLEX approximation. Note the absence of AL and particle-particle
exchange diagrams discussed in the text. (As before, the
coefficients $-1/2$ and $-3/2$ are omitted for clarity; see Equation~(\protect\ref{te28}).)}
\label{tf8}
\end{figure}

It is also important to re-emphasize at this point that the one-particle FLEX calculations
described here and in EB assume the exchange of elementary particle-hole fluctuations,
but not elementary particle-particle fluctuations. For this reason particle-particle
fluctuation propagators do not appear in crossed-channel contributions to $\Delta\Gamma_{\rm d}$
and $\Delta\Gamma_{\rm m}$ below. 
In analogy with Equations~(\ref{te5}) and (\ref{te6}) the spin-diagonalized particle-hole
vertices (see Figure~\ref{tf8})
may be written as follows:
\begin{eqnarray}
\label{te28}
 \gamph_{\rm d}(12;\,34)
&=&V_{\rm d}(12;\,34)\ -\ \half\Phi_{\rm d}(42;\,31)\ -\ \threehalf\Phi_{\rm m}
  (42;\,31) \\
\label{te29}
 \gamph_{\rm m}(12;\,34)
&=&V_{\rm m}(12;\,34)\ -\ \half\Phi_{\rm d}(42;\,31)\ +\ \half\Phi_{\rm m}
  (42;\,31)\ .
\end{eqnarray}
The functions $\Phi_{\rm d}$ and $\Phi_{\rm m}$ are as defined previously. 

In terms of the center-of-mass momentum-frequency $Q$ the density vertex takes the
form
\begin{eqnarray}
\label{te30}
\lefteqn{ \gamph_{\rm d}(Q;\,m_1m_2,\,\Delta{\bf R}_{12},\,i\omega;\,
  m_3m_4,\,\Delta{\bf R}_{34},\,i\omega')\ =}\nn
&& V_{\rm d}(\bfQ;\,m_1m_2,\,\Delta{\bf R}_{12};
  \,m_3m_4,\,\Delta{\bf R}_{34})\nn  
&&\mbox{}\ +\ \sum_{\Delta{\bf R}_{13}}\,e^{-i\bfQ\cdot\Delta{\bf R}_{13}}\,\Delta\gamph_{\rm d}
  (\Delta{\bf R}_{13},\,i\Omega;\,m_1m_2,\,\Delta{\bf R}_{12},\,i\omega;\,m_3m_4,\,\Delta{\bf R}_{34},\,
  i\omega')\ , \nn
\end{eqnarray}
with
\begin{eqnarray}
\label{te31}
\lefteqn{ \Delta\gamph_{\rm d}
  (\Delta{\bf R}_{13},\,i\Omega;\,m_1m_2,\,\Delta{\bf R}_{12},\,i\omega;\,m_3m_4,\,\Delta{\bf R}_{34},\,
  i\omega')\ =}\nn
  &&{1\over N}\sum_{\bfQ'}\,e^{i\bfQ'\cdot\Delta{\bf R}_{43}}\,
   \Bigl[-\half\Phi_{\rm d}-\threehalf\Phi_{\rm m}\Bigr](\bfQ',\,i(\omega'-\omega);\,m_4m_2,\, 
\Delta{\bf R}_{42};\,m_3m_1,\,\Delta{\bf R}_{31})\ ,\nn
\end{eqnarray}
where, as in Equation~(\ref{te23}), all relative displacements are expressed in terms of the set
$\{\Delta{\bf R}_{13},\,\Delta{\bf R}_{12},\,\Delta{\bf R}_{34}\}$. The analogous expression for
$\gamph_{\rm m}$ follows by the correspondence in Equation~(\ref{te29}). 

The particle-hole eigenvalue problem takes the form
\begin{equation}
\label{te32}
\gamph_{\rm r}(Q)\gph(Q)\phi(Q)\ =\ \lambda(Q)\phi(Q)\ , 
\end{equation}
where now
\begin{eqnarray}
\label{te33}
\lefteqn{ \gph(Q;\,m_1m_2,\,\Delta{\bf R}_{12},\,i\omega;\,m_3m_4,\,
  \Delta{\bf R}_{34},\,i\omega')\ =}\nn
  && \delta_{\omega\omega'}
  {T\over N}\sum_{\bfk}\,e^{i\bfk\cdot(\Delta{\bf R}_{12}-\Delta
  {\bf R}_{34})}\,G_{m_1m_3}(k+Q)G_{m_4m_2}(k)\ .
\end{eqnarray}
As in Equation~(\ref{te27}), a positive eigenvalue indicates attraction.

%% file: res.tex
\subsection{Sources of Systematic Error}
\label{ch4resa}

In this section we discuss the nature of the errors which arise in 
calculation of instability eigenvalues and transition temperatures.
Four main sources of error arise in the eigenvalue calculations.
These are accumulation of frequency-space renormalization
group\cite{tr20} error at low temperatures; error from the use of frequency cutoffs; 
error from truncation of the two-body vertex function in the 
relative real-space coordinate; and
$\bf k$-space discretization error ($16\times 16$ meshes are employed throughout).
Detailed discussions of the renormalization group procedure for the one-particle
self-energy are included in EB and Reference~\onlinecite{tr20}. The errors associated with this 
approximation are generally negligibly small in comparison with the other sources.

The frequency cutoff used in our calculations is $\Omega_c=0.5t_{\rm pd}$
for the ingoing and outgoing frequencies $\omega$ and $\omega'$ (see Figure~\ref{tf6})
in the fluctuation-induced component of the singlet kernel.
For the instantaneous part of the singlet kernel, whose decrease at high frequencies
is controlled solely by the falloff of the uncorrelated propagator $G^{\rm pp}$,
the corresponding cutoff is $50t_{\rm pd}$.  Errors associated with these cutoffs 
are extremely small.
For example, for the standard parameter set (see Equation~(\ref{te34})) at 16\% hole doping
and temperature $T=t_{\rm pd}/512$ (29 K), the $\dxy$ eigenvalue obtained using
the cutoffs described above is $\lambda_{\rm d}=1.0458$. If both cutoffs are raised to
$50t_{\rm pd}$, the eigenvalue becomes 1.0459, a change of 0.01\%; this demonstrates
the calculation's insensitivity to the cutoff associated with the fluctuation component.
In contrast, if both cutoffs are 
dropped to $0.5t_{\rm pd}$, the eigenvalue becomes 1.0439, a change of 0.2\%; this
demonstrates insensitivity to the cutoff associated with the instantaneous component.
It should be noted that at higher temperatures the cutoff on the fluctuation component
must be raised to obtain comparable percentage accuracy. This is
not costly, however, since the density of Matsubara frequencies decreases at the same time.

Next we discuss the truncation procedure for dealing with the
relative real-space coordinate in the two-body vertex. When the 
kernels are evaluated on a $16\times 16$ $\bf k$-space grid, the relative
displacements $\Delta\bfR_{12}$ and $\Delta\bfR_{34}$ (see Figure~\ref{tf6}) may take on
256 different values. Since the $\dxy$
eigenfunctions fall off rapidly at large values of $\Delta\bfR_{12}$ (see Section~\ref{ch4resd}),
it is rather intuitive to introduce a truncated basis set for the relative displacements.
In our calculations we limit the basis set to the twenty-one smallest
lattice vectors; i.e., elements of the kernel are zeroed out for $|\Delta\bfR|>a\sqrt{5}$.
The corresponding gain in computation time is approximately $(256/21)^2\sim 150$.

\begin{figure}[hbtp]
\begin{picture}(150,165)
\includegraphics{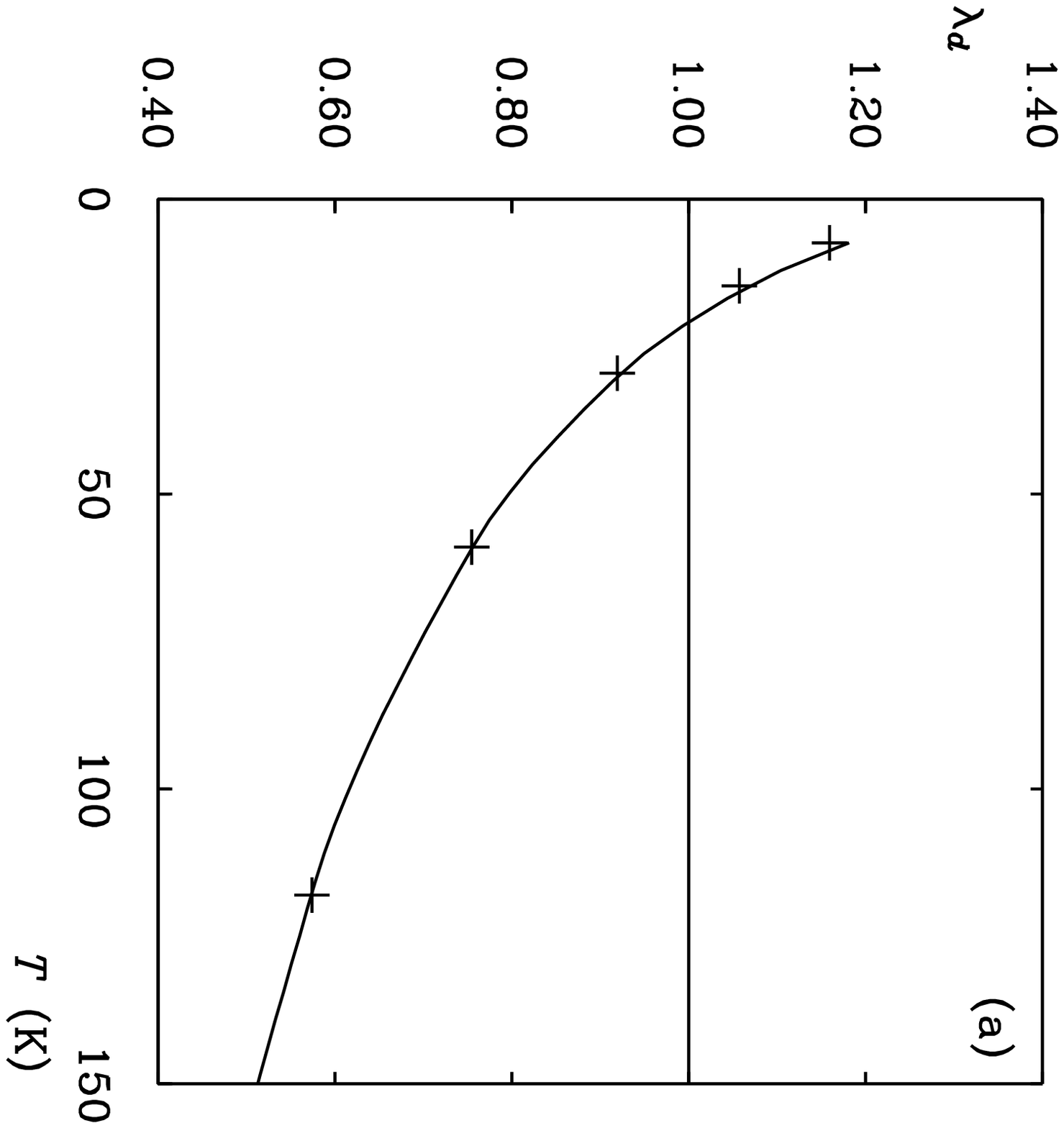}
\includegraphics{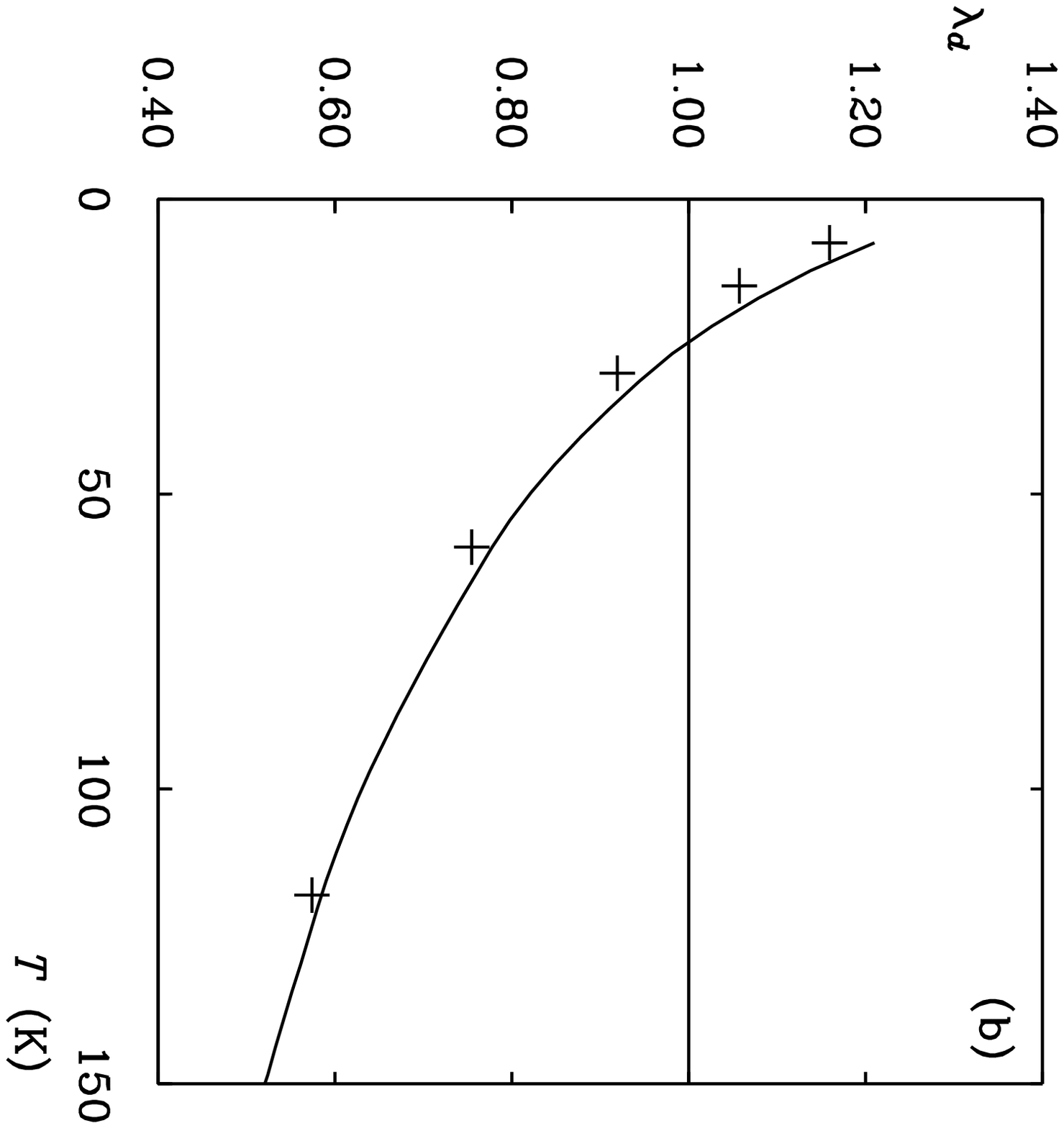}
\end{picture}
\caption{Systematic error analysis for eigenvalue and $T_c$ calculations in the
$\Udd$-only model. All parameters are at their standard values (Equation~(\protect\ref{te34}))
except that $\Upp=\Upd=0$.  The filling is $\navg=1.16$. 
(a)~Comparison of temperature-dependent $\dxy$ eigenvalues calculated using a full basis
of relative displacement states (solid line) and the 21-state basis with $|\DeltaR |\leq a
\protect\sqrt{5}$
(crosses). The $\bfk$-space mesh is $16\times 16$.
(b)~Comparison of eigenvalues calculated using a $32\times 32$ discretization (solid line)
and a $16\times 16$ discretization (crosses). The 21-state truncated basis is employed.}
\label{tf9}
\end{figure}

Since the calculation of the full model with the untruncated real-space
basis set is too time-consuming to be practical, we have used the simpler model
with $U_{\rm pp}=U_{\rm pd}=0$ for an error analysis. The behavior of the two models
is expected to be identical as far as this error check is concerned.
In Figure~\ref{tf9} we plot the temperature dependence of the $\dxy$ eigenvalue 
for the $U_{\rm dd}$-only model using the untruncated basis set and
the 21-state basis set. The difference in the eigenvalues
is very small for the two cases. For example, at $T=t_{\rm pd}/1024$ (15 K), 
$\lambda_{\rm d}=1.0683$ with the untruncated
basis and $\lambda_{\rm d}=1.0572$ with the 21-state basis. The corresponding
$T_c$ values are 20.8 K and 20.1 K, justifying the use of the truncated basis set.

The biggest source of error in the calculation of the instability eigenvalues is
the use of a $16\times 16$ $\bf k$-space discretization. For the models under study,
the low-temperature eigenvalues from a $16\times 16$ and a $32\times 32$ 
discretization differ by less than
5\%. This discretization error is very similar to that in previous studies of
the one-band Hubbard model.\cite{tr19,tr21}
This means one should also expect roughly the same size error (i.e., 5\%) in
comparing the $16\times 16$ results to the fine-mesh limit. 

In the figure below we plot the temperature dependence of the $\dxy$ eigenvalue 
for the $U_{\rm dd}$-only model using $16\times 16$ and $32\times 32$ 
discretizations. (Essentially identical behavior is expected for the 
the full CuO$_2$ model.) At $T=t_{\rm pd}/1024$ (15 K),
$\lambda_{\rm d}=1.1034$ for the $32\times 32$ study and 1.0572 for the
$16\times 16$ study.
For both cases a 21-state real-space basis truncation has been employed.
The corresponding $T_c$ values are 24 K and 20 K, corresponding to
an underestimation of $T_c$ by 4 K using the $16\times 16$ discretization. 

As mentioned in Section~\ref{ch4der} we have employed a Lanczos-Arnoldi algorithm\cite{tr16}
to calculate the first few maximum-real-part eigenvalues
in each scattering channel. This algorithm is especially
powerful for sparse matrices because it requires only repetitive
multiplication of a vector by the matrix of interest. Since a large fraction of
the elements in our scattering kernels are non-zero, but negligibly small with
regard to calculation of the large eigenvalues, a
sparse storage scheme is appropriate. For the scheme adopted throughout most of
our calculations, eigenvalues are affected by less than a few parts in a thousand, and
the gain in storage is of order 50.

\subsection{Eigenvalues for Particle-Particle Channels}
\label{ch4resb}

In the plots which follow we make use of a ``standard'' CuO$_2$ parameter set
derived for undoped La$_2$CuO$_4$ by Hybertsen, Schl$\rm\ddot u$ter, and Christensen\cite{tr9}
using constrained-occupancy density functional theory. These standard parameters for the
Hamiltonian in Equations~(\ref{te1}) and (\ref{te2}) are as follows:
\begin{eqnarray}
\label{te34}
            \tpd&\simeq&1.3\ {\rm eV}\  =\ 15,100\ {\rm K} \nn
            \tpp&\simeq&0.65\ {\rm eV}\ =\ 0.5\tpd \nn
            \eps&\simeq&3.6\ {\rm eV}\  =\ 2.75\tpd \nn
            \Udd&\simeq&10.5\ {\rm eV}\ =\ 8\tpd \nn
            \Upp&\simeq&4\ {\rm eV}\    =\ 3\tpd \nn
            \Upd&\simeq&1.2\ {\rm eV}\  =\ \tpd\ .  
\end{eqnarray}

\begin{figure}[hbtp]
\begin{picture}(150,85)
\includegraphics{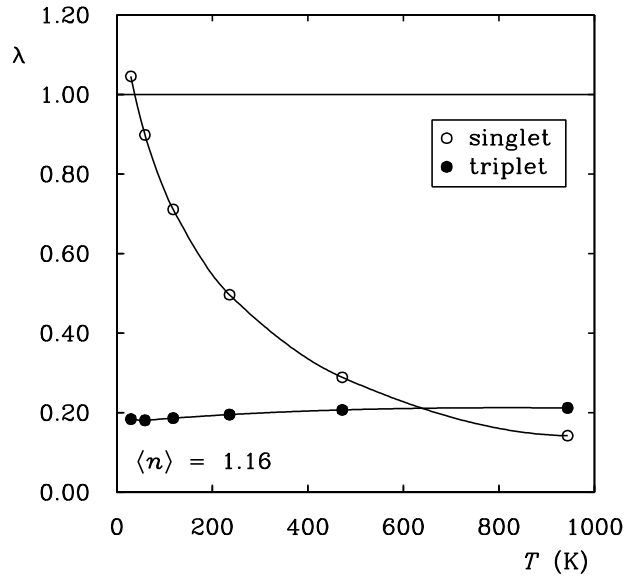}
\end{picture}
\caption{Temperature dependence of the maximal singlet and triplet eigenvalues
for the standard parameter set at $\navg=1.16$. The singlet eigenfunction has $\dxy$
symmetry, and the triplet state odd-frequency s-wave symmetry. The $\dxy$ eigenvalue reaches
unity, signaling a superconducting transition, at $T=37$ K.}
\label{tf10}
\end{figure}

\begin{figure}[hbtp]
\begin{picture}(80,75)
\includegraphics{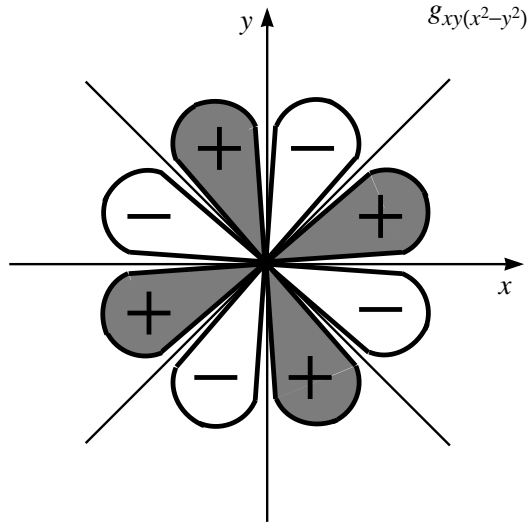}
\end{picture}
\caption{Schematic representation of the nodal structure of the 
g$_{\rm xy(x^2-y^2)}$ singlet state. Note that 
the subscript, when viewed as a function, vanishes on the locus of nodes (just as
in the case of the $\dxy$ state).}
\label{tf11}
\end{figure}

The temperature dependence of the maximal particle-particle eigenvalues for the standard
parameter set at $\navg=1.16$ (16\% hole doping) is illustrated in Figure~\ref{tf10}. 
The maximal singlet eigenvalue corresponds to a $\dxy$
state. This eigenvalue reaches unity, indicating a superconducting transition,
at $T/\tpd=0.0025$, i.e., $T=37$ K. At the transition temperature the next-leading
singlet eigenvalue is of order 0.4 and corresponds to a state with so-called
g-wave symmetry (i.e., nodes on the $x$ and $y$ axes, as well as the lines
$x=\pm y$; see Figure~\ref{tf11}). A third eigenvalue, corresponding to an orthogonal 
$\dxy$ state, lies just below the g-wave.

The maximal triplet channel eigenvalue in Figure~\ref{tf10} remains
small ($\sim 0.2$) throughout the temperature range of interest. The triplet state
in this case is antisymmetric in frequency and s-wave-like (i.e., symmetric) in space.
(Note in this regard that our instability analysis includes all eigenvectors of the scattering
kernels, including exotic singlet and triplet states with an antisymmetric frequency
dependence.)

\begin{figure}[hbtp]
\begin{picture}(150,85)
\includegraphics{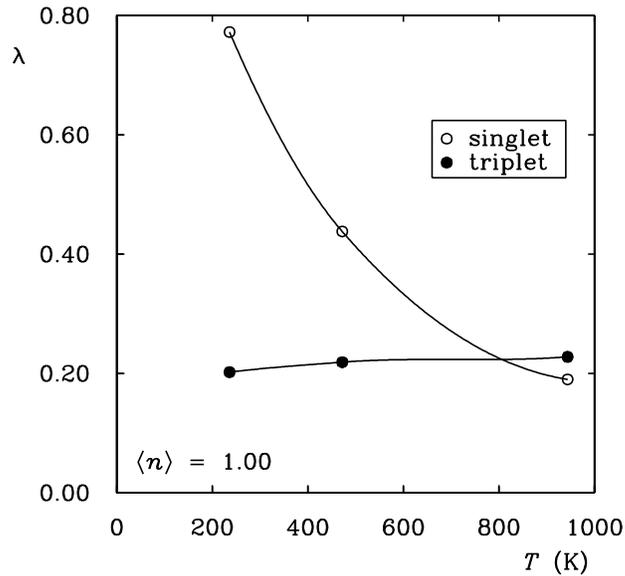}
\end{picture}
\caption{Temperature dependence of the maximal singlet and triplet
eigenvalues for the standard parameter set at $\navg=1.00$. The eigenfunction
symmetries are as in Figure~\protect\ref{tf10}.}
\label{tf12}
\end{figure}

For comparison the behavior of the maximal particle-particle eigenvalues at $\navg=1.00$
is illustrated in Figure~\ref{tf12}. The extreme singularity of the magnetic fluctuations in
this case prevents study at temperatures lower than $\tpd/64$, i.e., $T=240$ K.

\subsection{Transition Temperatures for $\dxy$ Superconductivity}
\label{ch4resc}

Eigenvalue plots of the type illustrated in Figure~\ref{tf10} may be used to
extract transition temperatures for the $\dxy$ singlet. The critical behavior of this FLEX
transition is classical, despite the fact that it is driven 
entirely by fluctuations. In terms of the FLEX eigenvalues,
\begin{equation}
\label{te35}
\lambda_{\rm d}(T)\ \sim\ 1-a(T-T_c) 
\end{equation}
for $T\sim T_c$, with $a>0$. This contrasts with the exact critical
behavior for a two-dimensional superconducting transition in the xy 
universality class:
\begin{equation}
\label{te36}
\lambda_{\rm xy}(T)\ \sim\ 1- B(T) e^{-A/\sqrt{T-T_{\rm xy}}}\ , 
\end{equation}
with $A$ a positive constant and $B(T)$ an algebraic function.\cite{tr22}
It is nevertheless possible to interpret the
$\dxy$ instability in FLEX as a ``mean-field'' transition with respect
to critical order parameter fluctuations.
With this caveat it is of interest to examine
the dependence of this instability on doping and model parameters.
[A more sophisticated treatment of the interference between the $\dxy$ transition and the
incipient instability in the magnetic channel is
presumably necessary for a detailed understanding of the pseudogap regime
observed in experiments,\cite{tr23} but that is not our intention here. In fact an additional
charge density state, the so-called ``orbital antiferromagnet''\cite{tr24}
or ``flux phase,''\cite{tr25} is also apparently relevant in the pseudogap regime; 
see the discussion of this state in Section~\ref{ch4rese}.]

In the plots which follow transition temperatures are
given in units of K; they may be rescaled in units of $\tpd$ using the correspondence 
in Equation~(\ref{te34}). The experimentally observed transition temperatures\cite{tr26} for
La$_{2-\rm x}$Sr$_{\rm x}$CuO$_4$ are plotted for comparison using
the assumed correspondence
\begin{equation}
\label{te37}
{\rm x}\ \rightarrow\ \navg-1\ . 
\end{equation}

\begin{figure}[hbtp]
\begin{picture}(150,85)
\includegraphics{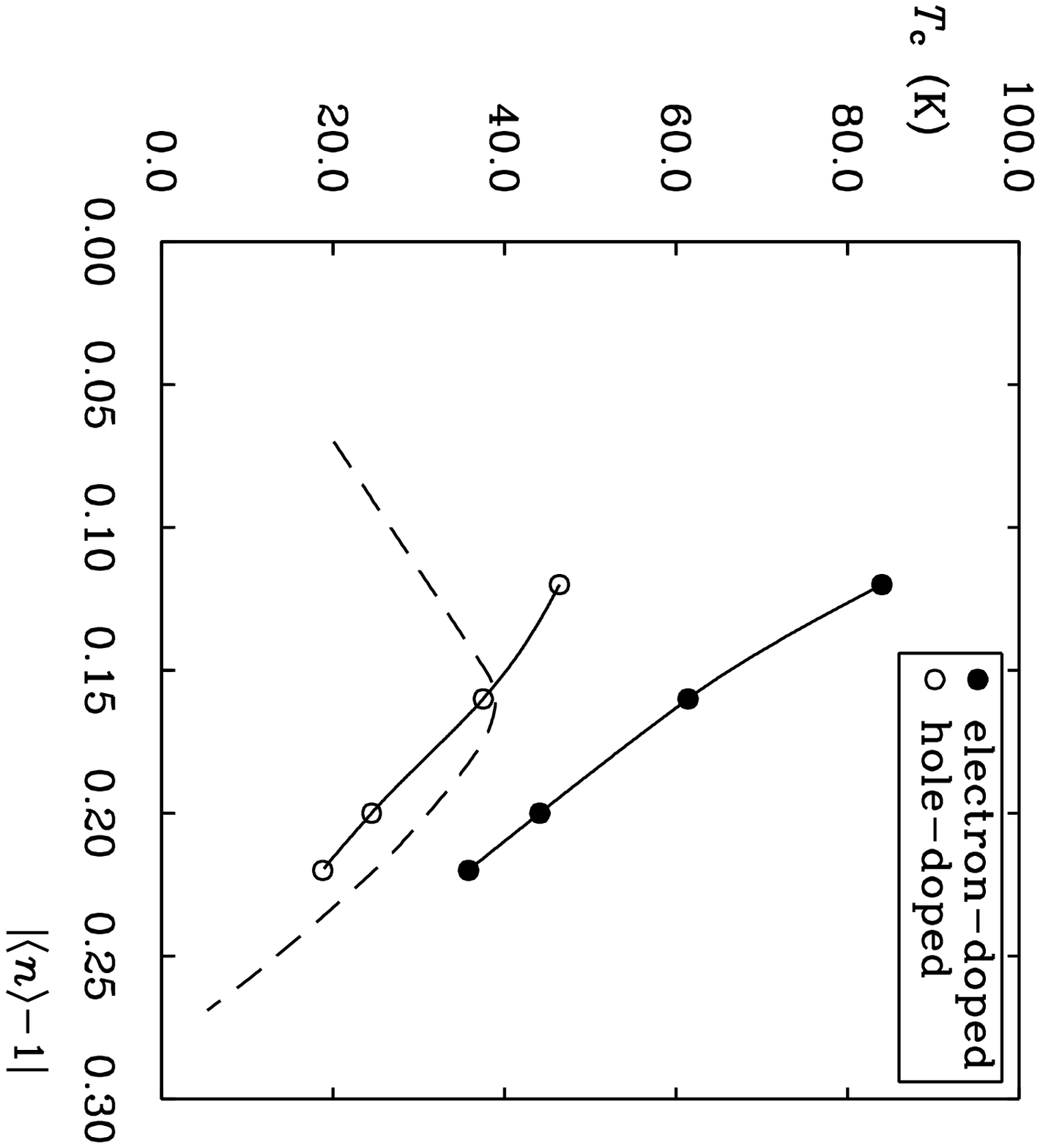}
\end{picture}
\caption{Doping dependence of the $\dxy$ transition for the standard
parameter set. Results are shown for both $\navg>1.00$ (hole doping) and $\navg<1.00$
electron doping. For comparison the doping dependence of the experimental transition~\protect\cite{tr26}
in La$_{2-\rm x}$Sr$_{\rm x}$CuO$_4$ is plotted (dashed line) using the assumed correspondence
${\rm x}\rightarrow \navg-1$. The increase of the FLEX $T_c$ on the electron-doped side
is largely due to an increase in the spin fluctuation strength.}
\label{tf13}
\end{figure}

As shown in Figure~\ref{tf13}, a $\dxy$ transition occurs for both hole doping
($\navg$ greater than 1) and electron doping ($\navg$ less than 1).
Since the CuO$_2$ model has only approximate particle-hole symmetry around the
point $\navg=1$, the transition temperatures are not symmetric. Within our
FLEX calculation the pairing interaction becomes increasingly singular
as $\navg\rightarrow 1$, and we have only been able to calculate superconducting
instability
temperatures for doping levels greater than 12\%. (In any case a self-consistent
parquet-like treatment of vertex functions\cite{tr14,tr19} seems essential for values of $\navg$
closer to unity.) The higher transitions
for electron doping are consistent with the presence of enhanced
magnetic fluctuations on this side of the phase diagram.\cite{tr12,tr13}
The transition temperatures on the hole-doped side are
strikingly similar to the experimental curve in the overdoped regime,
$\navg-1 > 0.16$. At smaller doping the FLEX curve continues to rise, while the
experimental curve peaks and turns down in the underdoped region. As
remarked previously, in this region the $\dxy$ singlet channel is in
strong competition with the $\bfQ\sim (\pi,\,\pi)$ antiferromagnetic spin
channel, as well as an exotic $\bfQ\sim (\pi,\,\pi)$ charge
density channel (see also Section~\ref{ch4rese}). It is tempting to speculate that the 
downturn in the
experimental $\dxy$ transition temperature results from this competition.

In the next six figures we examine the sensitivity of the $\dxy$
transition temperature to changes in the model parameters. Our discussion
is limited to the hole-doped side of the phase diagram. First we alter 
a single parameter at a time, keeping other parameters fixed at their
standard values, then we briefly consider the behavior of the
drastically simplified CuO$_2$ model with $\Upp=\Upd=0$. 

\begin{figure}[t]
\begin{picture}(150,85)
\includegraphics{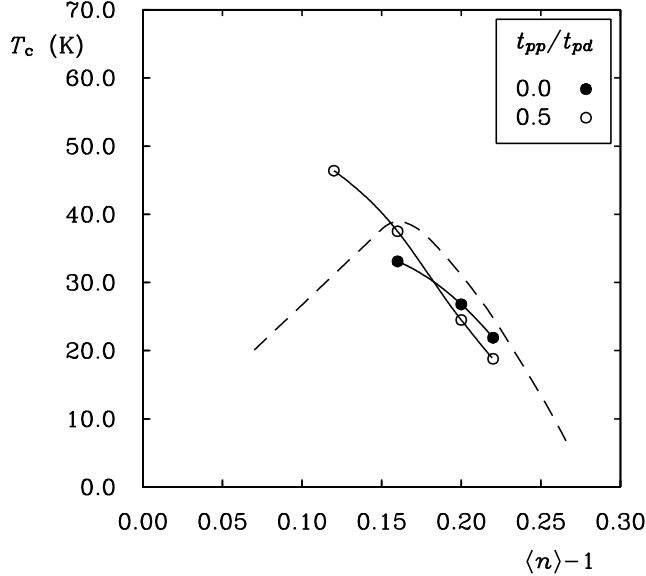}
\end{picture}
\caption{Dependence of $T_c$ on the O--O hopping integral $\tpp$.
In Figures~\protect\ref{tf14}--\protect\ref{tf19}
all model parameters are set to their standard values except as noted. Further the
experimental curve~\protect\cite{tr26} for doped lanthanum cuprate (dashed line) is superimposed 
for comparison.}
\label{tf14}
\end{figure}

The effect of removing the O--O hopping integral $\tpp$ is shown in
Figure~\ref{tf14}. This change alters the shape of the
Fermi surface,\cite{tr12} improving the degree of nesting and enhancing the
spin fluctuation spectrum. However, the transition temperature remains essentially unchanged,
since the positive effect on the singlet vertex
is largely compensated by a reduction in the uncorrelated propagator $G^{\rm pp}$.

\begin{figure}[hbtp]
\begin{picture}(150,85)
\includegraphics{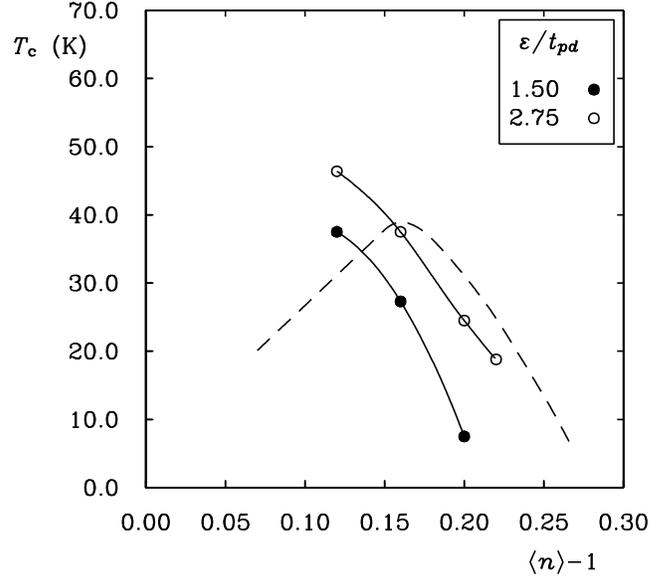}
\end{picture}
\caption{Dependence of $T_c$ on the unrenormalized Cu--O level separation
$\varepsilon=\varepsilon_{\rm p}-\varepsilon_{\rm d}$. The behavior of the model for
$\varepsilon=0$ was also examined, but no transition occurs in this case.}
\label{tf15}
\end{figure}

The effect of changing the Cu--O orbital separation $\varepsilon=\varepsilon_{\rm p}-\varepsilon_{\rm d}$ is much more drastic, as expected. The value of
$\varepsilon$ largely determines the strength of the spin fluctuations.
(This is because $\varepsilon$ is smaller than $\Udd$, i.e., the system is
in the so-called charge-transfer regime.\cite{tr7}) For small values of $\varepsilon$,
occupation of the O orbitals becomes comparable to occupation of the Cu
orbitals (or even larger, when Coulomb interactions are taken into account). As an example,
for $\varepsilon=0$ and $\navg=1.16$, only 33\% of the holes reside on the Cu orbitals.
Since $\Upp$ is considerably less than $\Udd$, increased O occupancy
reduces the strength of the spin fluctuation propagator and weakens the
pairing tendency. This fact is illustrated clearly in Figure~\ref{tf15}. The $\dxy$
transition temperature drops sharply when $\varepsilon$ is reduced from
3.6 eV to 2.0 eV. The transition disappears completely when $\varepsilon$ is
set to zero (i.e., the bare Cu and O orbitals become degenerate); this is due not only
to the reduction of the effective Coulomb parameter, but also to the loss of nesting in
the $\varepsilon=0$ Fermi surface.

The dependence of $T_c$ on the Coulomb parameters $\Udd$, $\Upp$, and
$\Upd$  is relatively complex, since these parameters contribute to
both the one- and two-body effective interactions. For the simpler
one-band Hubbard model\cite{tr21} (and for the CuO$_2$ model with $\Udd$ only---see the discussion below), an increase of the Coulomb integral leads to
a peak $T_c$, then a gradual decrease at larger values. The origin of this
behavior is a competition between the pairing vertex (which is
enhanced by a large Coulomb interaction) and the uncorrelated propagator $G^{\rm pp}$
(which is suppressed). In these simpler models the on-site Coulomb
interaction does not {\sl directly} suppress pairing, since the $\dxy$
state has no on-site pairing component. While this remains true for
$\Udd$ in the full CuO$_2$ model, it is not necessarily
true for $\Upp$ and $\Upd$: the $\dxy$ pair wave function generally has on-site
O--O and near-neighbor Cu--O components, which are suppressed by the
Coulomb integrals $\Upp$ and $\Upd$. The importance of this direct
effect depends on the admixture of the relevant components in the
$\dxy$ pair state (see the discussion of the pair wave function in Section~\ref{ch4resd}).

For the O--O Coulomb integral $\Upp$, this direct suppression of pairing
apparently dominates, i.e., an increase in $\Upp$ leads to more repulsion
in the $\dxy$ pair state and a reduced transition temperature (Figure~\ref{tf16}).
As discussed in Section~\ref{ch4resd} below, the $\dxy$ pair does have a non-zero
on-site O--O component, consistent with the observed trend in
$T_c$.
\begin{figure}[hbtp]
\begin{picture}(150,85)
\includegraphics{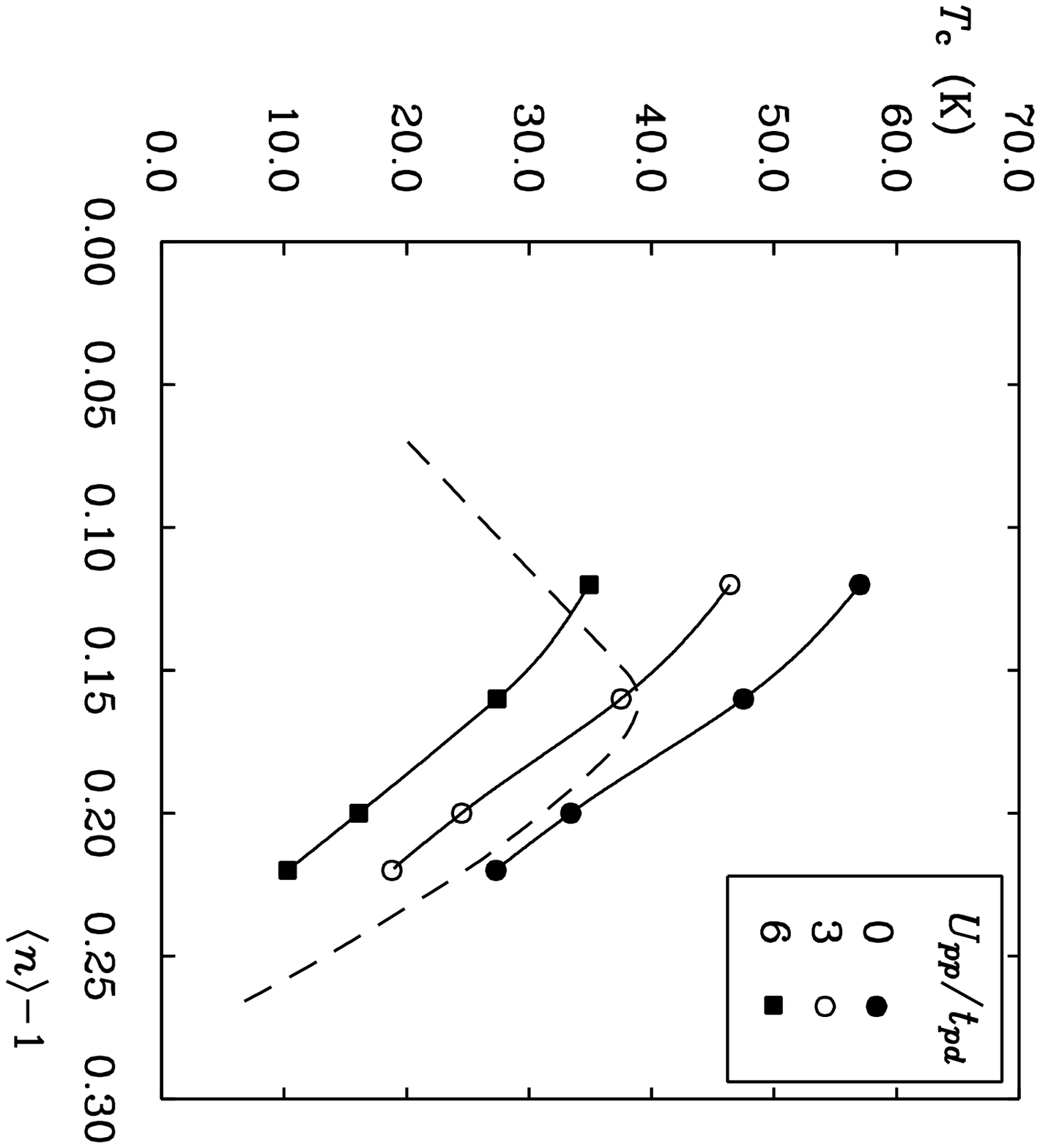}
\end{picture}
\caption{Dependence of $T_c$ on the intra-orbital O--O Coulomb integral $\Upp$.}
\label{tf16}
\end{figure}

\begin{figure}[hbtp]
\begin{picture}(150,85)
\includegraphics{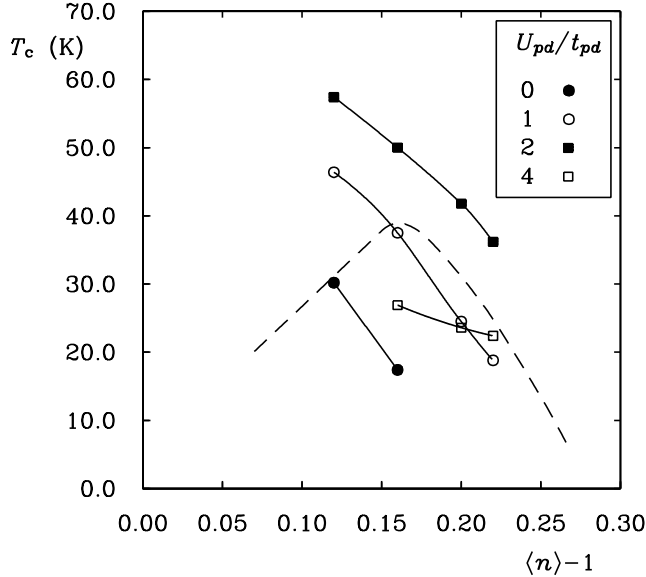}
\end{picture}
\caption{Dependence of $T_c$ on the near-neighbor Cu--O Coulomb integral
$\Upd$.}
\label{tf17}
\end{figure}

For the Cu--O Coulomb integral $\Upd$, the trend in $T_c$ with increasing
$\Upd$ (Figure~\ref{tf17}) resembles the trend with increasing $U$ in the one-band Hubbard model:
An increase, peak, and subsequent decrease.
This behavior reflects a compromise between
the enhancement of the spin fluctuations and pairing interaction with increased
$\Upd$; the accompanying suppression of $G^{\rm pp}$; and the direct suppression of
$T_c$ noted above. The enhancement of the spin fluctuations with increasing $\Upd$ arises
from improved Fermi surface nesting and from increased d-orbital occupancy due to 
a rise in the Hartree-Fock level separation $\varepsilon_{\rm\scriptscriptstyle HF}$ 
(see the discussion of the Hartree-Fock Hamiltonian in EB).
[It is interesting to note that the direct suppression of $T_c$ is 
moderated by the same phenomenon which reduces the effective Coulomb repulsion in
conventional electron-phonon superconductivity:\cite{tr27} The near-neighbor component of the
pair wave function changes sign at high frequencies, effectively reducing the repulsion
in the low-frequency region, i.e., inducing a Coulomb pseudopotential.]

\begin{figure}[hbtp]
\begin{picture}(150,85)
\includegraphics{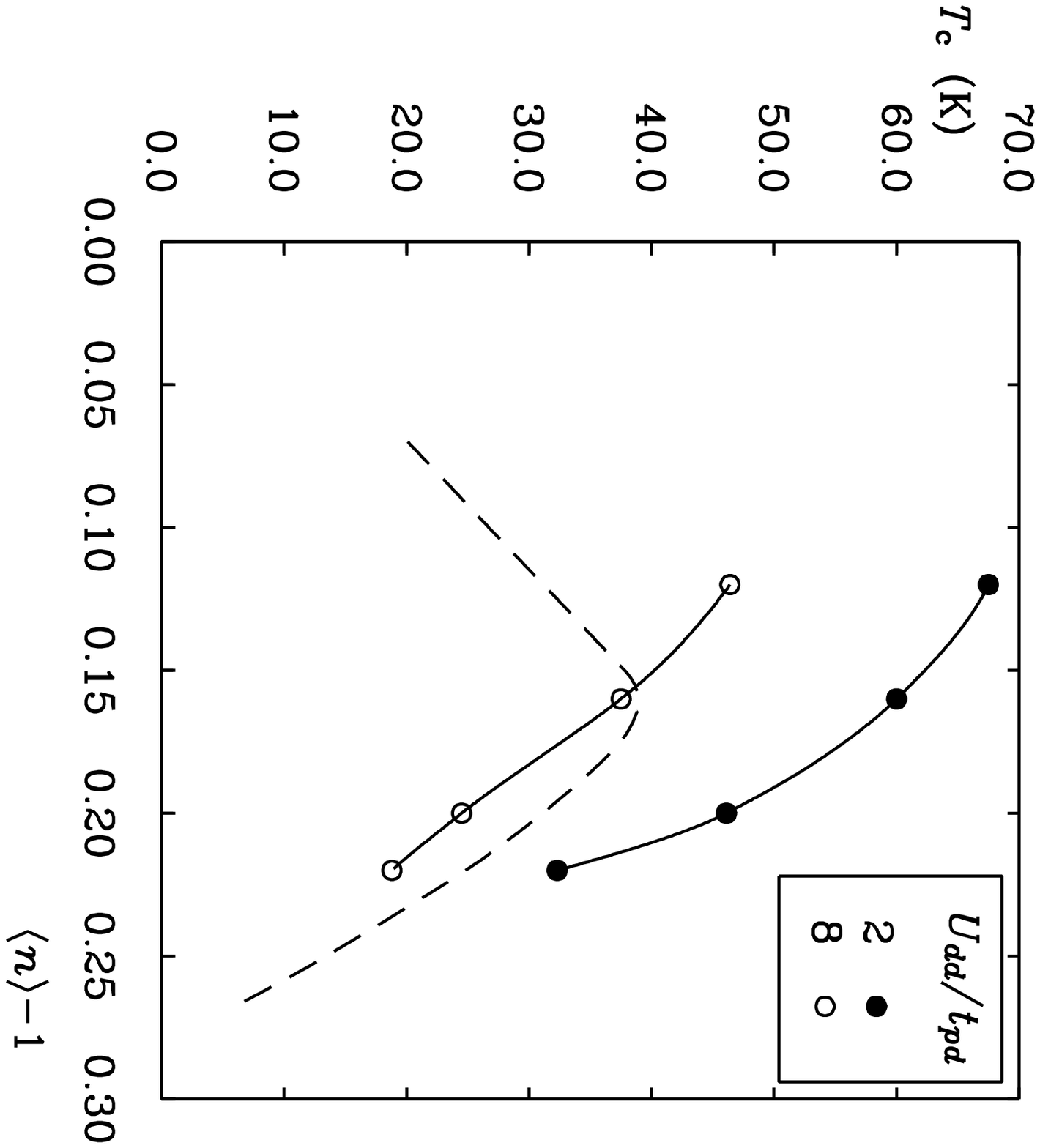}
\end{picture}
\caption{Dependence of $T_c$ on the intra-orbital Cu--Cu Coulomb integral
$\Udd$.}
\label{tf18}
\end{figure}

Finally Figure~\ref{tf18} shows the transition temperatures for two different
values of $\Udd$, 10.5 eV (standard parameter set---charge-transfer regime)
and 2.5 eV (Hubbard regime).
The substantial reduction in $\Udd$ in this case
has only a minimal effect on the strength of the spin fluctuations.\cite{tr12}
This is because a {\sl decrease} in $\Udd$ results in an increased Hartree-Fock level 
separation 
$\varepsilon_{\rm\scriptscriptstyle HF}$ and an increased d-orbital occupancy (cf. the
discussion of an {\sl increase} in $\Upd$ above). The increased d-orbital occupancy
offsets the direct effect of a smaller $\Udd$ in the spin fluctuation propagator. 
The increase in $T_c$ for this admittedly unrealistic 
parameter set then results from an increase in $G^{\rm pp}$ (i.e., a density of states effect).

The CuO$_2$ model with $\Upp=\Upd=0$ (the ``$\Udd$-only model'') has
been studied previously\cite{tr13} at temperatures above the $\dxy$ transition.
This model is conceptually problematic: the omission of the interactions associated with
the p-orbitals substantially alters the Hartree-Fock Fermi surface, largely 
negating any improvement in
the band structure expected from the addition of the extra bands. (Both $G^{\rm pp}$ and
the d-orbital spin fluctuation strength are significantly affected by the omission.)
Furthermore, while the $\dxy$ wave function is dominated by d-orbital components, the
omission of $\Upp$ and $\Upd$ completely eliminates those components associated with the
p-orbitals. The principal virtue of the $\Udd$-only model
is its calculational simplicity. Since the Coulomb interaction $\Udd$ is
zero-range, the computations involved are essentially the same as those
in the one-band Hubbard model. For example, the particle-hole ladders
$\Phi$ in Section~\ref{ch4der} become scalar, rather than matrix, inverses.

\begin{figure}[hbtp]
\begin{picture}(150,165)
\includegraphics{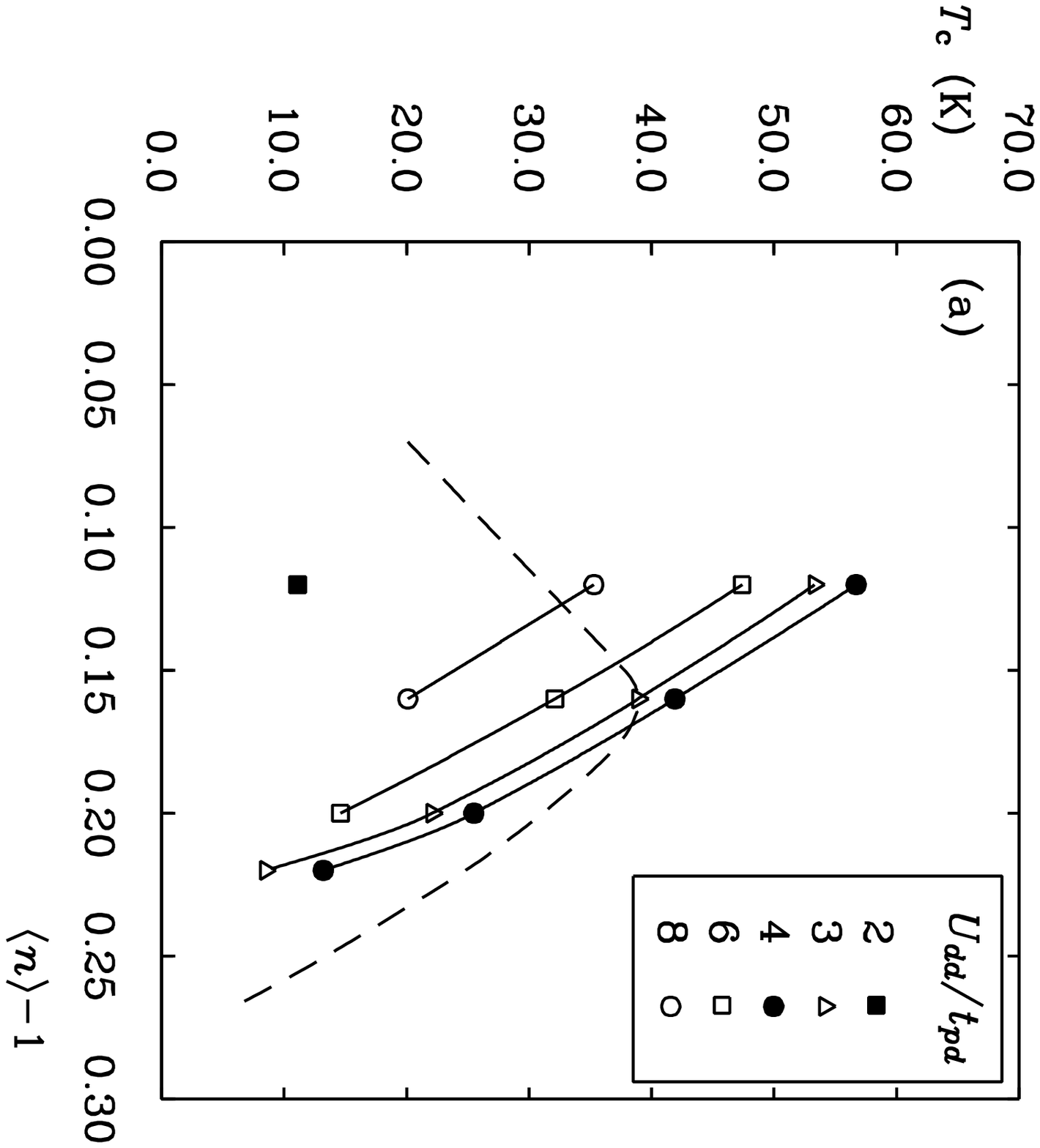}
\includegraphics{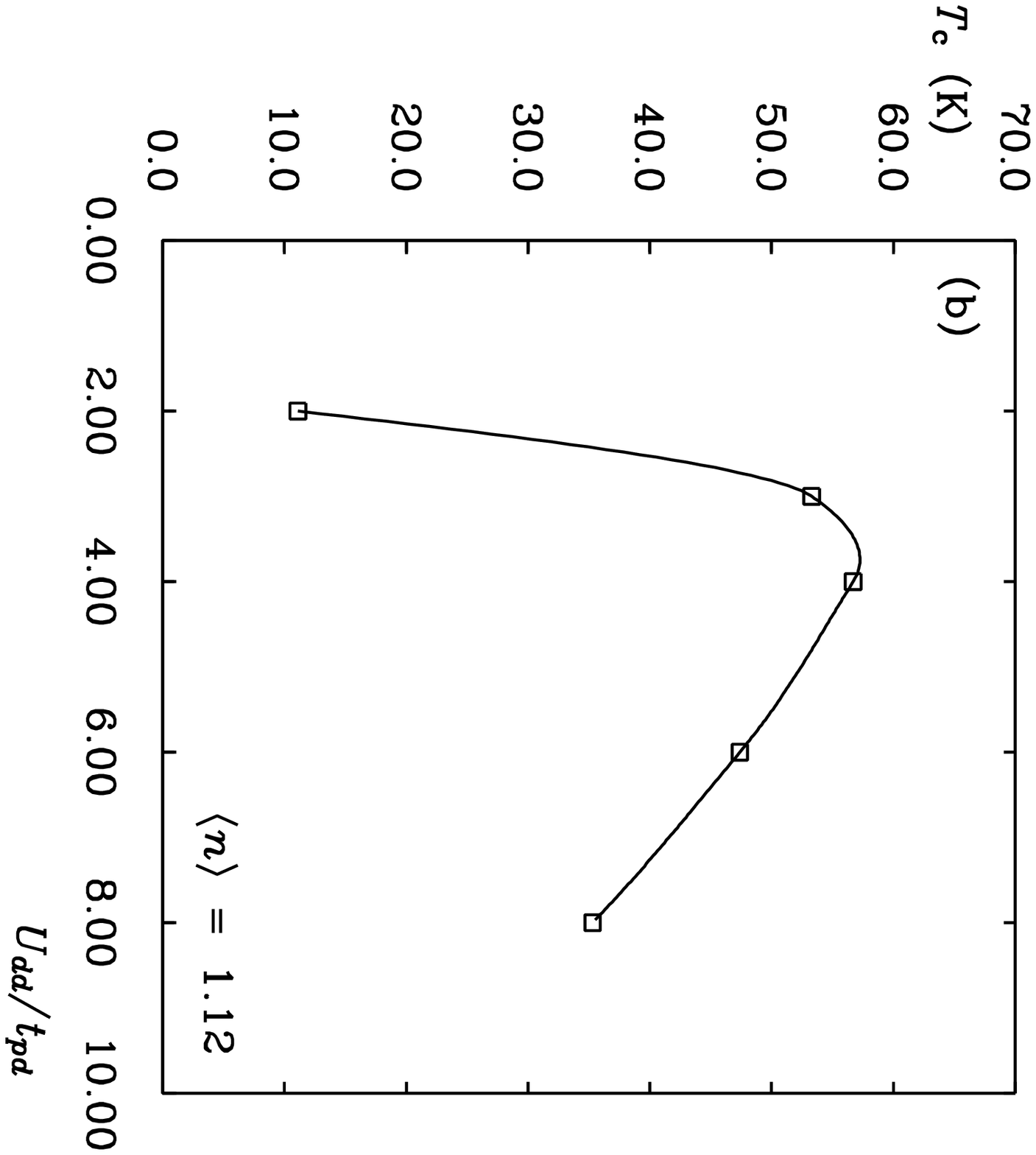}
\end{picture}
\caption{Dependence of $T_c$ on the intra-orbital Cu--Cu Coulomb
integral $\Udd$ in the ``$\Udd$-only'' model. The Coulomb integrals $\Upp$ and $\Upd$
are in this case set to zero, with other parameters remaining at their standard values.
(a)~Variation of $T_c$ with doping for several values of $\Udd$. The behavior for
$\Udd/\tpd=1$ was also examined, but no transition occurs in this case.
(b)~Variation of $T_c$ with $\Udd$ at fixed filling $\navg=1.12$.}
\label{tf19}
\end{figure}

For completeness, the variation of $T_c$ with $\Udd$ in the $\Udd$-only
model is shown in Figure~\ref{tf19}. As expected, the qualitative dependence of
$T_c$ is the same as that in the one-band Hubbard model:\cite{tr21} The peak value
of $T_c$ for increasing $\Udd$ is determined by a competition between
enhancement of the pairing interaction and suppression of the uncorrelated
propagator $G^{\rm pp}$.

\subsection{$\dxy$ Singlet Wave Function}
\label{ch4resd}

The graphical representation of the particle-particle pair eigenfunction
$\phi(Q;\,ab,$ $\Delta{\bf R}_{ab},\,i\omega)$ is hindered by its large number
of degrees of freedom. It is essential to make use of symmetries to
emphasize the eigenfunction's key features. The first basic symmetry
follows from the Pauli Principle. Written in terms of anticommuting
c-numbers, the pair state corresponding to eigenfunction $\phi_{\rm r}$ is
\begin{eqnarray}
\label{te38}
\lefteqn{ \bsp \sum_{\bfR;\,ab,\,\Delta\bfR_{ab},\,\omega}\!\!\!\!\!\!\!\! 
  e^{i\bfQ\cdot(\bfR+\Delta
  \bfR_{ab})}\,\phi_{\rm r}(Q;\,ab,\,\Delta\bfR_{ab},\,i\omega)}\nn
&&  \times\sum_{\sigma\sigma'}\,\chi_{\rm r}^{\sigma\sigma'}\,\cbar_{a\sigma}
  (\bfR+\Delta\bfR_{ab},\,i(\omega+\Omega))\cbar_{b\sigma'}(\bfR,\,
  -i\omega)\ , 
\end{eqnarray}
where r=s or t, with
\begin{equation}
\label{te39}
  \chi_{\rm s}\ =\ {1\over\sqrt 2}\left( \matrix{ 0 & 1 \cr -1 & 0 \cr}
  \right) 
\end{equation}
and
\begin{equation}
\label{te40}
  \chi_{\rm t}\ =\ \left( \matrix{ 1 & 0 \cr 0 & 0 \cr}\right),\ \ {1\over
   \sqrt 2}\left( \matrix{ 0 & 1 \cr 1 & 0 \cr}\right),\ \ {\rm or}
  \ \left( \matrix{ 0 & 0 \cr 0 & 1 \cr}\right)\ . 
\end{equation}
Since
\begin{equation}
\label{te41}
 \cbar_{a\sigma}\cbar_{b\sigma'}\ =\ -\cbar_{b\sigma'}\cbar_{a\sigma}\ ,
\end{equation}
it follows that
\begin{eqnarray}
\label{te42}
 \lefteqn{ \bsp \sum_{\bfR;\,ab,\,\Delta\bfR_{ab},\,\omega}\!\!\!\!\!\!\!\!
  e^{i\bfQ\cdot(\bfR+\Delta
  \bfR_{ab})}\,\phi_{\rm r}(Q;\,ab,\,\Delta\bfR_{ab},\,i\omega)}\nn
&&  \times\sum_{\sigma\sigma'}\,\chi_{\rm r}^{\sigma'\sigma}\,\cbar_{b\sigma}
  (\bfR,\,-i\omega)
  \cbar_{a\sigma'}(\bfR+\Delta\bfR_{ab},\,i(\omega+\Omega))\ =\nn
 \lefteqn{\bsp \sum_{\bfR;\,ab,\,\Delta\bfR_{ab},
  \,\omega}\!\!\!\!\!\!\!\!
  e^{i\bfQ\cdot(\bfR+\Delta
  \bfR_{ab})}\,\phi_{\rm r}(Q;\,ab,\,\Delta\bfR_{ab},\,i\omega)}\nn
&&  \times\sum_{\sigma\sigma'}
  \,\chi_{\rm r}^{\sigma\sigma'}\,\cbar_{a\sigma}
  (\bfR+\Delta\bfR_{ab},\,i(\omega+\Omega))\cbar_{b\sigma'}(\bfR,\,
  -i\omega)\ . 
\end{eqnarray}
It is convenient to relabel the dummy sums on the left by first
interchanging orbital indices $a$ and $b$, then letting
\begin{eqnarray}
\label{te43}
 \Delta\bfR_{ba}&=&-\Delta\bfR_{ab} \nn
  \bfR&=&\bfR'+\Delta\bfR_{ab} \nn
  -i\omega&=&i(\omega'+\Omega)\ .  
\end{eqnarray}
Dropping the primes on the dummy variables $\bfR'$ and $\omega'$ gives
\begin{eqnarray}
\label{te44}
\lefteqn{\bsp \sum_{\bfR;\,ab,\,\Delta\bfR_{ab},\,\omega}\!\!\!\!\!\!\!\!
  e^{i\bfQ\cdot\bfR}\,\phi_{\rm r}(Q;\,ba,\,-\Delta\bfR_{ab},\,-i(\omega
  +\Omega))}\nn
  &&\times\sum_{\sigma\sigma'}\,\chi_{\rm r}^{\sigma'\sigma}\,\cbar_{a\sigma}
  (\bfR+\Delta\bfR_{ab},\,i(\omega+\Omega))
  \cbar_{b\sigma'}(\bfR,\,-i\omega)\ =\nn
\lefteqn{\bsp - \bsp\!\! \sum_{\bfR;\,ab,\,\Delta\bfR_{ab},
  \,\omega}\!\!\!\!\!\!\!\!
  e^{i\bfQ\cdot(\bfR+\Delta
  \bfR_{ab})}\,\phi_{\rm r}(Q;\,ab,\,\Delta\bfR_{ab},\,i\omega)}\nn
&&\times\sum_{\sigma\sigma'}\,\chi_{\rm r}^{\sigma\sigma'}\,\cbar_{a\sigma}
  (\bfR+\Delta\bfR_{ab},\,i(\omega+\Omega))\cbar_{b\sigma'}(\bfR,\,-i\omega)\ . 
\end{eqnarray}
The Pauli symmetry relations follow by identifying coefficients:
\begin{equation}
\label{te45}
\phi_{\rm r}(Q;\,ba,\,-\Delta\bfR_{ab},\,-i(\omega+\Omega))\chi_{\rm r}
  ^{\sigma'\sigma}\ =\ -e^{i\bfQ\cdot\Delta\bfR_{ab}}\phi_{\rm r}(Q;\,
  ab,\,\Delta\bfR_{ab},\,i\omega)\chi_{\rm r}^{\sigma\sigma'}\ , 
\end{equation}
i.e.,
\begin{equation}
\label{te46}
\phi_{\rm s}(Q;\,ab,\,\Delta\bfR_{ab},\,i\omega)\ =\ e^{-i\bfQ\cdot
  \Delta\bfR_{ab}}\phi_{\rm s}(Q;\,ba,\,-\Delta\bfR_{ab},\,-i(\omega+
  \Omega)) 
\end{equation}
for the singlet channel, and
\begin{equation}
\label{te47}
\phi_{\rm t}(Q;\,ab,\,\Delta\bfR_{ab},\,i\omega)\ =\ -e^{-i\bfQ\cdot
  \Delta\bfR_{ab}}\phi_{\rm t}(Q;\,ba,\,-\Delta\bfR_{ab},\,-i(\omega+
  \Omega)) 
\end{equation}
for the triplet channel.
These symmetry relations assume a particularly simple form for the
case of interest here, $Q=0$. 

To aid in the graphical display of the pair wave function it is useful
to introduce a basis of one-particle states with simple transformation
properties under point group operations. While the Cu 3$\dxy$ orbital
transforms into itself under all symmetry operations, the O 2p$_{\rm x}$
and 2p$_{\rm y}$ orbitals are generally mixed. It is, however, possible
to form linear combinations of the p$_{\rm x}$ and p$_{\rm y}$ orbitals
which transform in simple ways. To derive the transformed orbitals
we rewrite the pair wave function $\phi$ in Equation~(\ref{te38}), holding the
coordinates of particle $a$ fixed. For brevity the spin and frequency
dependence of $\phi$ is temporarily suppressed. The
components of the wave function for $b=x$ and $y$ take the form
\begin{eqnarray}
\label{te48}
\lefteqn{\bsp\bsp\sum_{\bfR;\,a,\,\Delta\bfR_{ab},\,\omega}\!\!\!\!\!\!\!\! 
  e^{i\bfQ\cdot(\bfR+\Delta\bfR_{ab})}\,\cbar_a(\bfR+\Delta\bfR_{ab})}\nn
&&  \times \Bigl[\phi{\rm_r}(Q;\,ax,\,\Delta\bfR_{ab})\cbar_x(\bfR)
  \,+\,\phi_{\rm r}(Q;\,ay,\,\Delta\bfR_{ab})\cbar_y(\bfR)\Bigr]\ .    
\end{eqnarray}
In the first term the sums on $\bfR$ and $\Delta\bfR_{ab}$ may be shifted by
\begin{eqnarray}
\label{te49}
 \bfR&\rightarrow&\bfR-\widehat x \nn
            \Delta\bfR_{ab}&\rightarrow&\Delta\bfR_{ab}+\widehat x\ . 
\end{eqnarray}
A similar shift may be performed in the second term with $\widehat x\rightarrow
  \widehat y$. This results in an equivalent symmetrized version of the
wave function,
\begin{eqnarray}
\label{te50}
\lefteqn{\bsp\bsp\half\sum_{\bfR;\,a,\,\Delta\bfR_{ab},\,\omega}
  \!\!\!\!\!\!\!\! 
  e^{i\bfQ\cdot(\bfR+\Delta\bfR_{ab})}\,\cbar_a(\bfR+\Delta\bfR_{ab})}\nn
&&  \times \Bigl[\phi{\rm_r}(Q;\,ax,\,\Delta\bfR_{ab})\cbar_x(\bfR)
  \,+\,\phi_{\rm r}(Q;\,ax,\,\Delta\bfR_{ab}+\widehat x)\cbar_x(\bfR-\widehat
  x) \nn
  &&\!\mbox{}+\ \phi_{\rm r}(Q;\,ay,\,\Delta\bfR_{ab})
  \cbar_y(\bfR)\,+\,\phi_{\rm r}(Q;\,ay,\,\Delta\bfR_{ab}+\widehat y)
  \cbar_y(\bfR-\widehat y)\Bigr]  
\end{eqnarray}
The four c-numbers $\cbar_x(\bfR)$, $\cbar_x(\bfR-\widehat x)$,
$\cbar_y(\bfR)$, $\cbar_y(\bfR-\widehat y)$ may now be re-expressed
in terms of the linear combinations
\begin{equation}
\label{te51}
\left[ \matrix{ \cbar_D(\bfR) \cr \cbar_S(\bfR) \cr \cbar_X(\bfR)
  \cr \cbar_Y(\bfR) \cr}\right] \ =\ \left[ \matrix{ 1/2 & 1/2 &
  1/2 & 1/2 \cr 1/2 & 1/2 & -1/2 & -1/2 \cr 1/\sqrt{2} &
  -1/\sqrt{2} & 0 & 0 \cr 0 & 0 & -1/\sqrt{2} & 1/\sqrt{2} \cr} \right]
  \,\left[ \matrix{ \cbar_x(\bfR) \cr \cbar_x(\bfR-\widehat x) \cr
  \cbar_y(\bfR) \cr \cbar_y(\bfR-\widehat y) \cr} \right]\ . 
\end{equation}

\begin{figure}[hbtp]
\begin{picture}(150,100)
\includegraphics{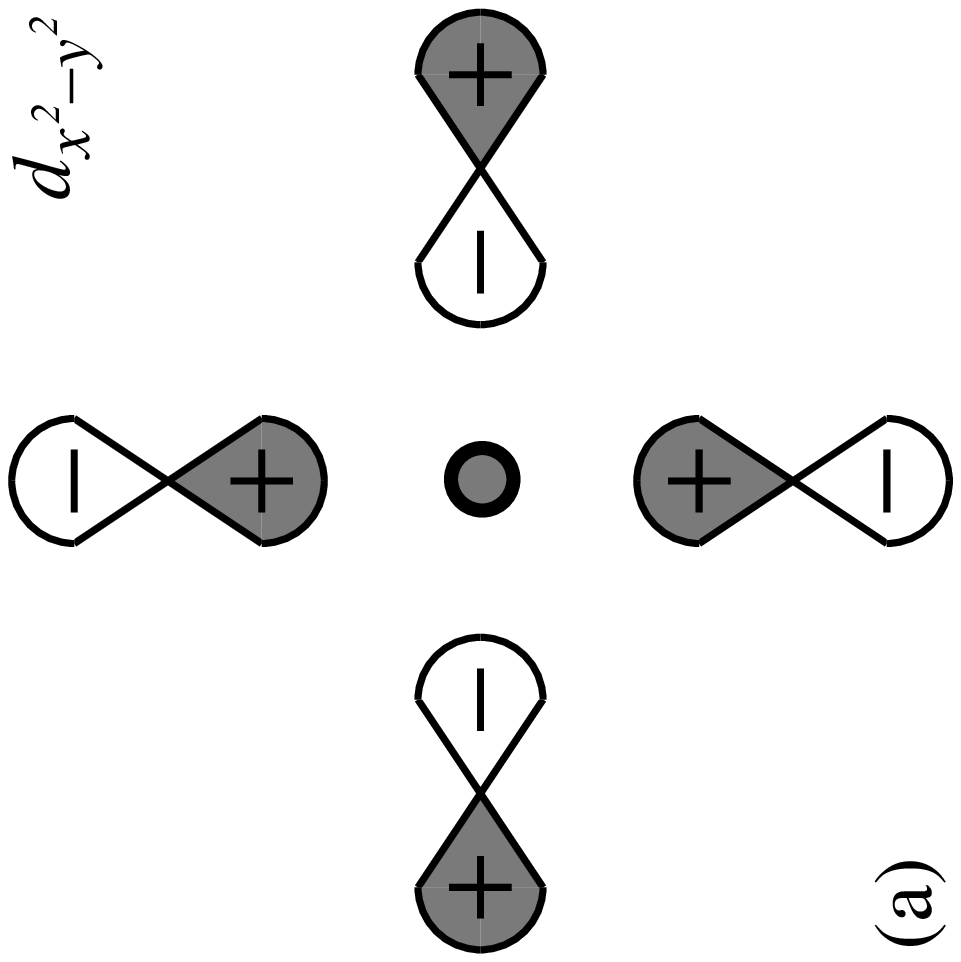}
\includegraphics{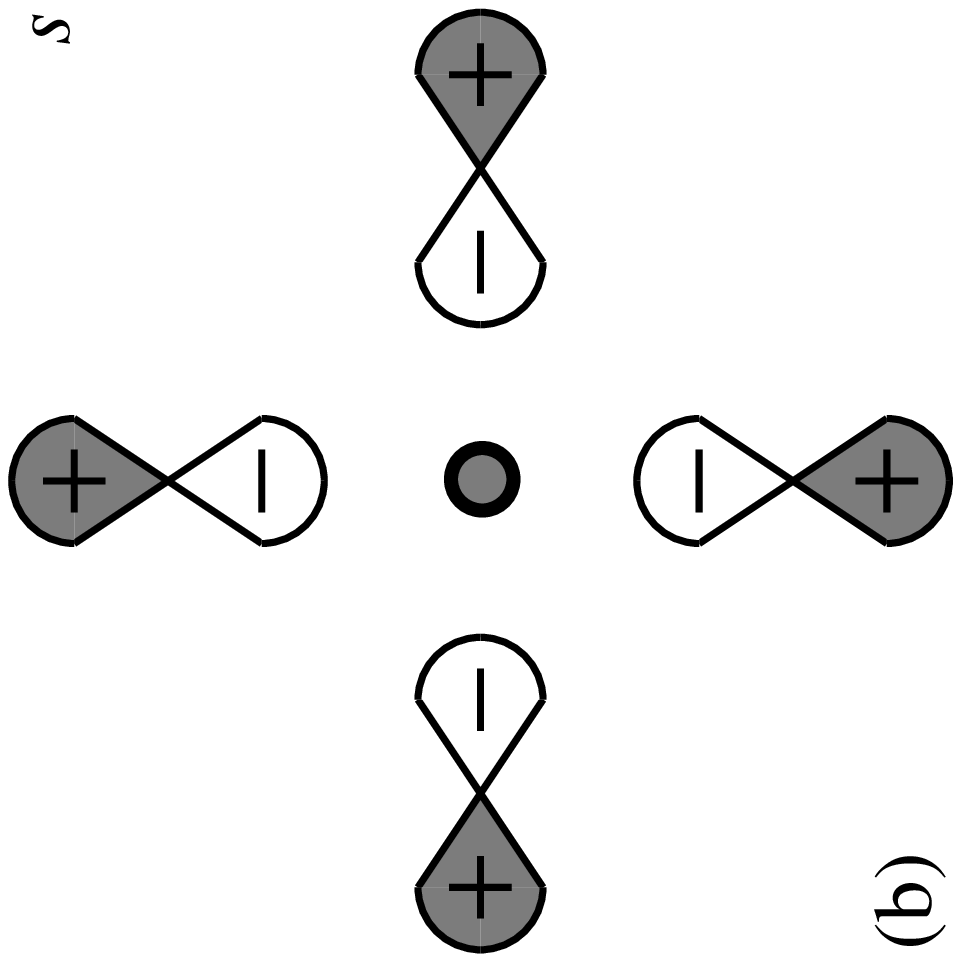}
\includegraphics{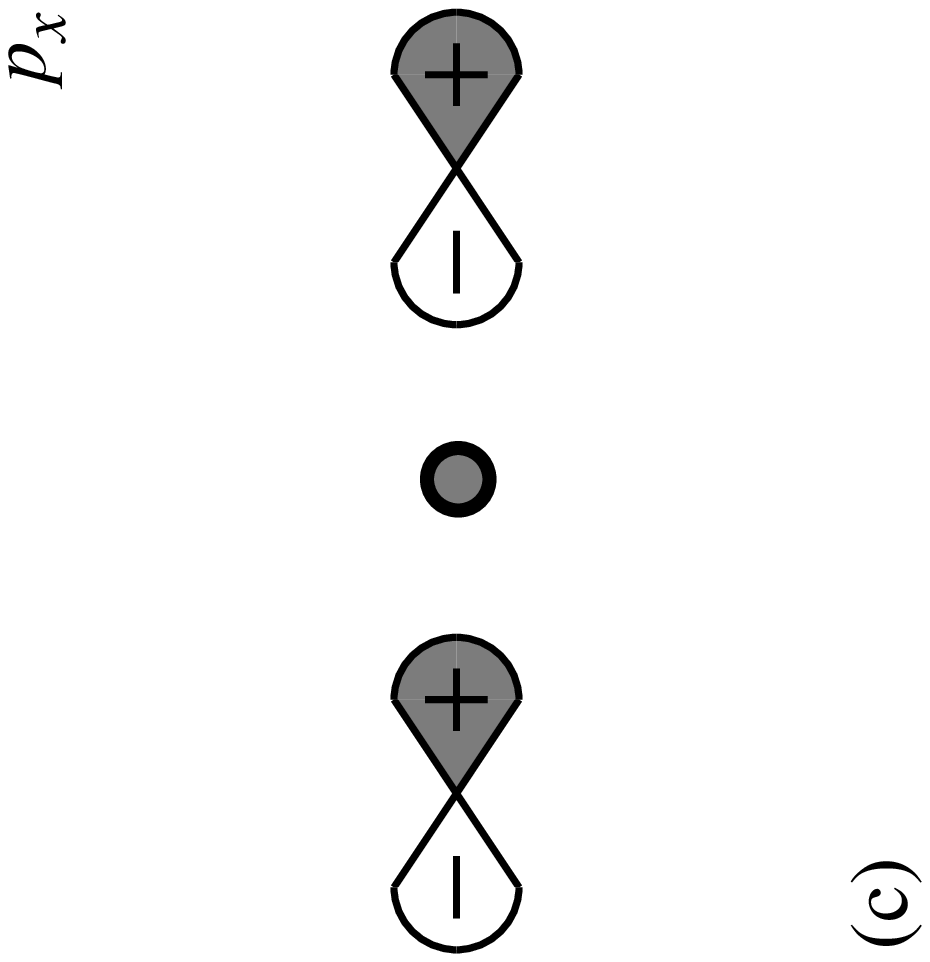}
\includegraphics{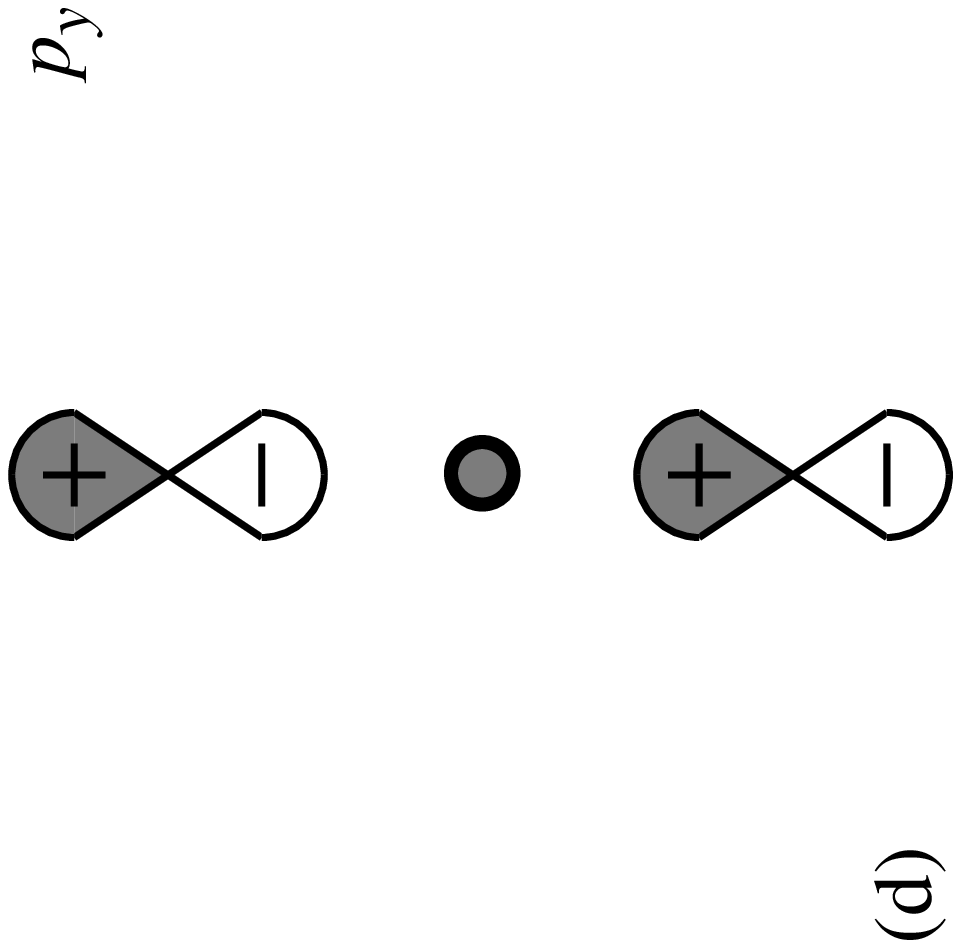}
\end{picture}
\caption{Extended p-orbital basis set with well-defined rotational
symmetry. The central Cu site is denoted by a shaded circle. (a)~Extended $\dxy$ 
orbital (state $D$).
(b)~Extended s orbital (state $S$). (c)~Extended p$_{\rm x}$ orbital (state $X$).
(d)~Extended p$_{\rm y}$ orbital (state $Y$).}
\label{tf20}
\end{figure}

The c-number $\cbar_D(\bfR)$ represents an extended oxygen orbital
with d$_{\rm x^2-y^2}$ rotational symmetry (Figure~\ref{tf20}(a)), just like the
central Cu 3d$_{\rm x^2-y^2}$ orbital. (The uniform phases in the
linear combination result from the initial definition of the 2p$_{\rm x}$
and 2p$_{\rm y}$ orbitals.) Likewise $\cbar_S(\bfR)$ represents an
extended s-wave oxygen orbital (Figure~\ref{tf20}(b)), and $\cbar_X(\bfR)$ and
$\cbar_Y(\bfR)$ represent extended p-wave orbitals (Figures~\ref{tf20}(c) and (d)).
The wave function components in Equation~(\ref{te50}) may now be rewritten
in terms of the new c-numbers. The complete pair wave function in
Equation~(\ref{te38}) then becomes
\begin{eqnarray}
\label{te52}
\lefteqn{\bsp\bsp\sum_{\bfR;\,aB,\,\Delta\bfR_{ab},\,\omega}\!\!\!\!\!\!\!\! 
  e^{i\bfQ\cdot(\bfR+\Delta
  \bfR_{ab})}\,\phi_{\rm r}(Q;\,aB,\,\Delta\bfR_{ab},\,i\omega)}\nn
  &&\times\sum_{\sigma\sigma'}\,\chi_{\rm r}^{\sigma\sigma'}\,\cbar_{a\sigma}
  (\bfR+\Delta\bfR_{ab},\,i(\omega+\Omega))\,\cbar_{B\sigma'}(\bfR,\,
  -i\omega)\ , 
\end{eqnarray}
where, as before, the sum on $a$ runs over $\{d,\,x,\,y\}$, but now the
sum on $B$ runs over $\{d,\,D,\,S,\,X,\,Y\}$. The new
wave function components are
\begin{eqnarray}
\label{te53}
\lefteqn{ \phi_{\rm r}(Q;\,aD,\,\Delta\bfR_{ab},\,i\omega)\ =} \nn
&&    \fourth\Bigl[\,\phi_{\rm r}(Q;\,ax,\,\Delta\bfR_{ab},\,i\omega)\,+\,\phi_{\rm r}
  (Q;\,ax,\,\Delta\bfR_{ab}+\widehat x,\,i\omega) \nn
&&\!\!\!\!\mbox{}\ +\ \phi_{\rm r}(Q;\,ay,\,\Delta\bfR_{ab},\,i\omega)\,+\,\phi_{\rm r}
  (Q;\,ay,\,\Delta\bfR_{ab}+\widehat y,\,i\omega)\,\Bigr] \nn 
 \lefteqn{ \phi_{\rm r}(Q;\,aS,\,\Delta\bfR_{ab},\,i\omega)\ =} \nn
  &&  \fourth\Bigl[\,\phi_{\rm r}(Q;\,ax,\,\Delta\bfR_{ab},\,i\omega)\,+\,\phi_{\rm r}
  (Q;\,ax,\,\Delta\bfR_{ab}+\widehat x,\,i\omega) \nn
&&\!\!\!\!\mbox{}\ -\ \phi_{\rm r}(Q;\,ay,\,\Delta\bfR_{ab},\,i\omega)\,-\,\phi_{\rm r}
  (Q;\,ay,\,\Delta\bfR_{ab}+\widehat y,\,i\omega)\,\Bigr] \nn 
 \lefteqn{\phi_{\rm r}(Q;\,aX,\,\Delta\bfR_{ab},\,i\omega)\ =} \nn
  &&  \halfsqrthalf\Bigl[\,\phi_{\rm r}(Q;\,ax,\,\Delta\bfR_{ab},\,i\omega)
  \,-\,\phi_{\rm r}(Q;\,ax,\,\Delta\bfR_{ab}+\widehat x,\,i\omega)\,\Bigr] \nn
  \lefteqn{ \phi_{\rm r}(Q;\,aY,\,\Delta\bfR_{ab},\,i\omega)\ =} \nn
&& \halfsqrthalf\Bigl[\,-\phi_{\rm r}(Q;\,ay,\,\Delta\bfR_{ab},\,i\omega)
  \,+\,\phi_{\rm r}(Q;\,ay,\,\Delta\bfR_{ab}+\widehat y,\,i\omega)\,\Bigr]\ . 
\end{eqnarray}
Note that this expression for the pair wave function is equivalent to
that in Equation~(\ref{te38}). The new basis for the $B$-particles is
simply overcomplete; the c-numbers $\cbar_{B\sigma'}(\bfR,\,i\omega)$
and $\cbar_{B\sigma'}(\bfR',\,i\omega)$ for near-neighbor $\bfR$ and
$\bfR'$ are no longer independent.

The new basis is well-adapted for representing pair wave functions
for $\bfQ=0$ in a simple way. It is convenient to keep the $B$-particle orbital
and unit-cell position fixed while varying $a$ and $\Delta\bfR_{ab}$.
For example, separate plots describe the system for the $B$-particle in
the Cu 3$\dxy$ orbital ($\phi_{ad}$), in the extended oxygen $\dxy$
orbital ($\phi_{aD}$), and in the extended oxygen s orbital
($\phi_{aS}$). If the pair wave function is to have overall $\dxy$
symmetry, $\phi_{ad}$ and $\phi_{aD}$ must have {\sl explicit}
$\dxy$ symmetry in the variable $\Delta\bfR_{ab}$, while $\phi_{aS}$
must have explicit s symmetry. (States with d symmetry may also be
constructed for $b$-particles in the $X$ and $Y$ orbitals. These states are
more complicated, since both $\phi_{aX}$ and $\phi_{aY}$ must be 
non-zero.)

\begin{figure}[hbtp]
\begin{picture}(150,105)
\includegraphics{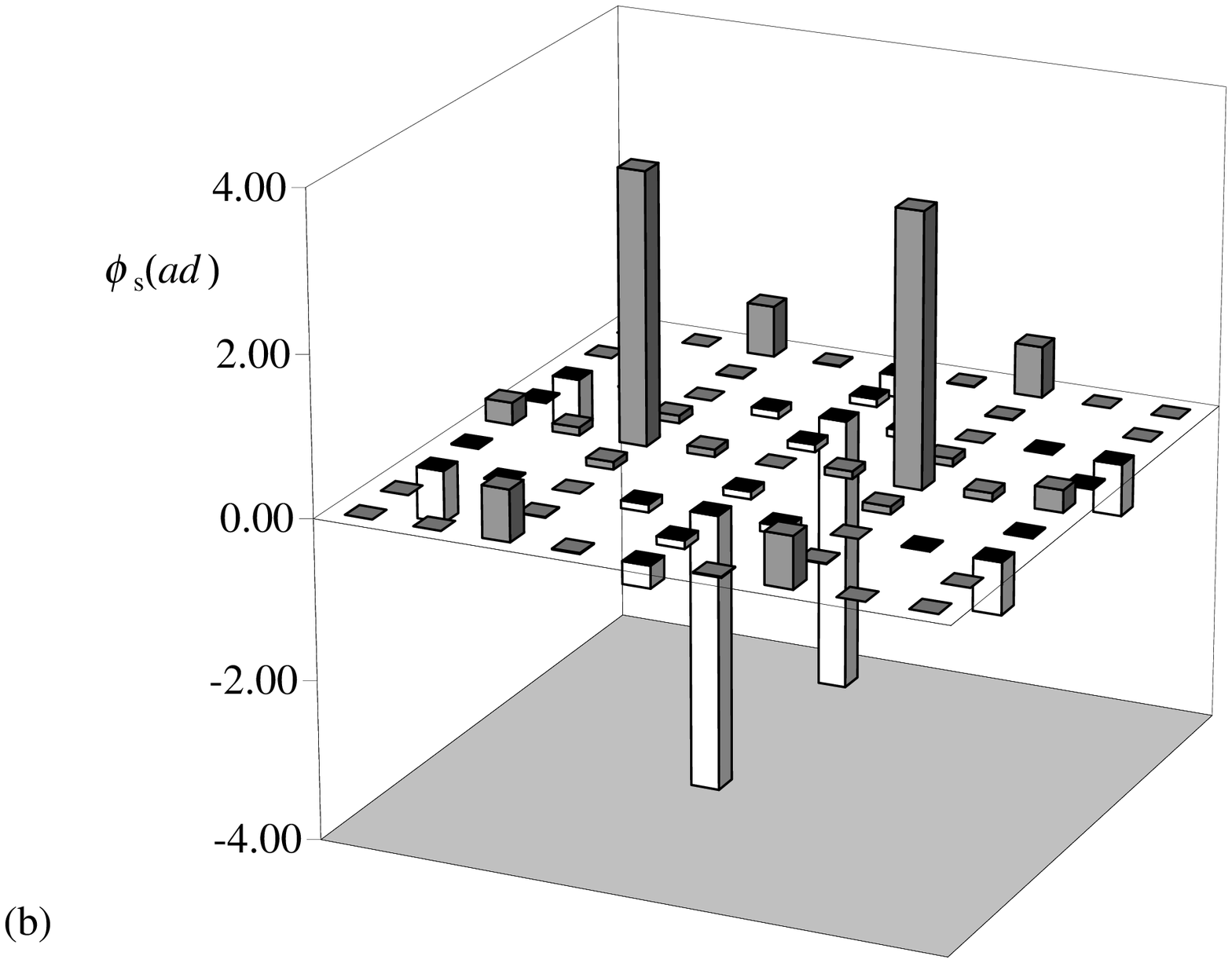}
\includegraphics{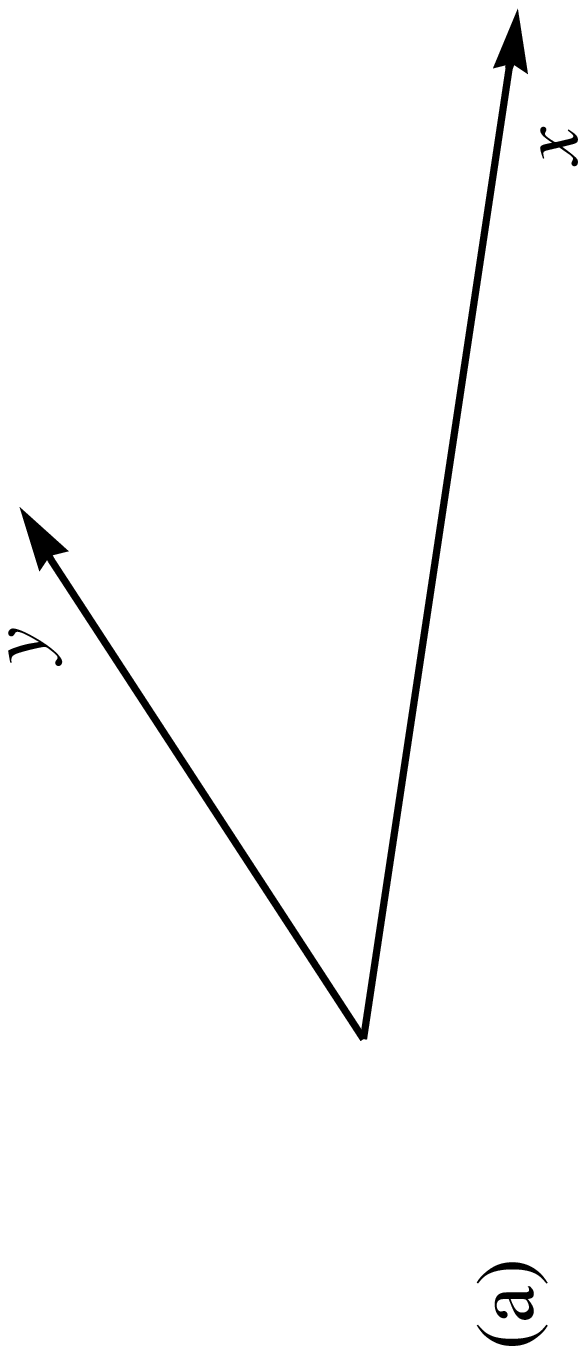}
\end{picture}
\caption{Relative displacement dependence of the minimum-frequency 
($\omega=\pi T$) components of the $\dxy$ pair wave function
at $T=\tpd/512=29$ K for the standard parameter set at $\navg=1.16$. The histograms are centered
on a Cu site in the cell with $\DeltaR=0$. Note that the wave function is real-valued.
(a)~Orientation of the $x$ and $y$ axes in the histogram plots.
(b)~Component $\phi_{\rm s}({\bf 0};\,ad,\DeltaR,\,i\pi T)$. 
($\DeltaR$ labels the unit-cell displacement to orbital $a$ from a 
fixed Cu 3d$_{\rm x^2-y^2}$ orbital.) Note that the histogram
exhibits explicit $\dxy$ symmetry in this case.
(c)~Component $\phi_{\rm s}({\bf 0};\,aD,\DeltaR,\,i\pi T)$. In this case the fixed
orbital is the extended O 2p linear combination with $\dxy$ symmetry. 
As before the histogram exhibits
explicit $\dxy$ symmetry. Note the difference in scale between this plot and that in (b).
(d)~Component $\phi_{\rm s}({\bf 0};\,aS,\DeltaR,\,i\pi T)$. In this case the fixed
orbital is the extended O 2p linear combination with s-wave symmetry. 
In this case the histogram exhibits explicit s-wave symmetry (see the discussion in the text).}
\label{tf21}
\end{figure}

In Figure~\ref{tf21} we use histograms to show the spatial variation of the
minimum-frequency ($\omega=\pi T$) components of the $\dxy$ singlet wave function 
(i.e., the right eigenvector of the
particle-particle kernel) for $T\sim T_c$. Recall that this function evolves
smoothly into the off-diagonal pair field below $T_c$. Each histogram shows a $4\times 4$
patch of unit cells, centered on a Cu site in the cell at $\Delta\bfR_{ab}=0$. The
orientation of the $x$ and $y$ axes is indicated in Figure~\ref{tf21}(a). The height of
the block at each point in the lattice is just the value of $\phi$ 
at that combination of $a$-particle orbital and displacement indices $(a,\,\Delta\bfR_{ab})$ 
for fixed $B$-particle indices.
\begin{figure}[hbtp]
\begin{picture}(150,230)
\includegraphics{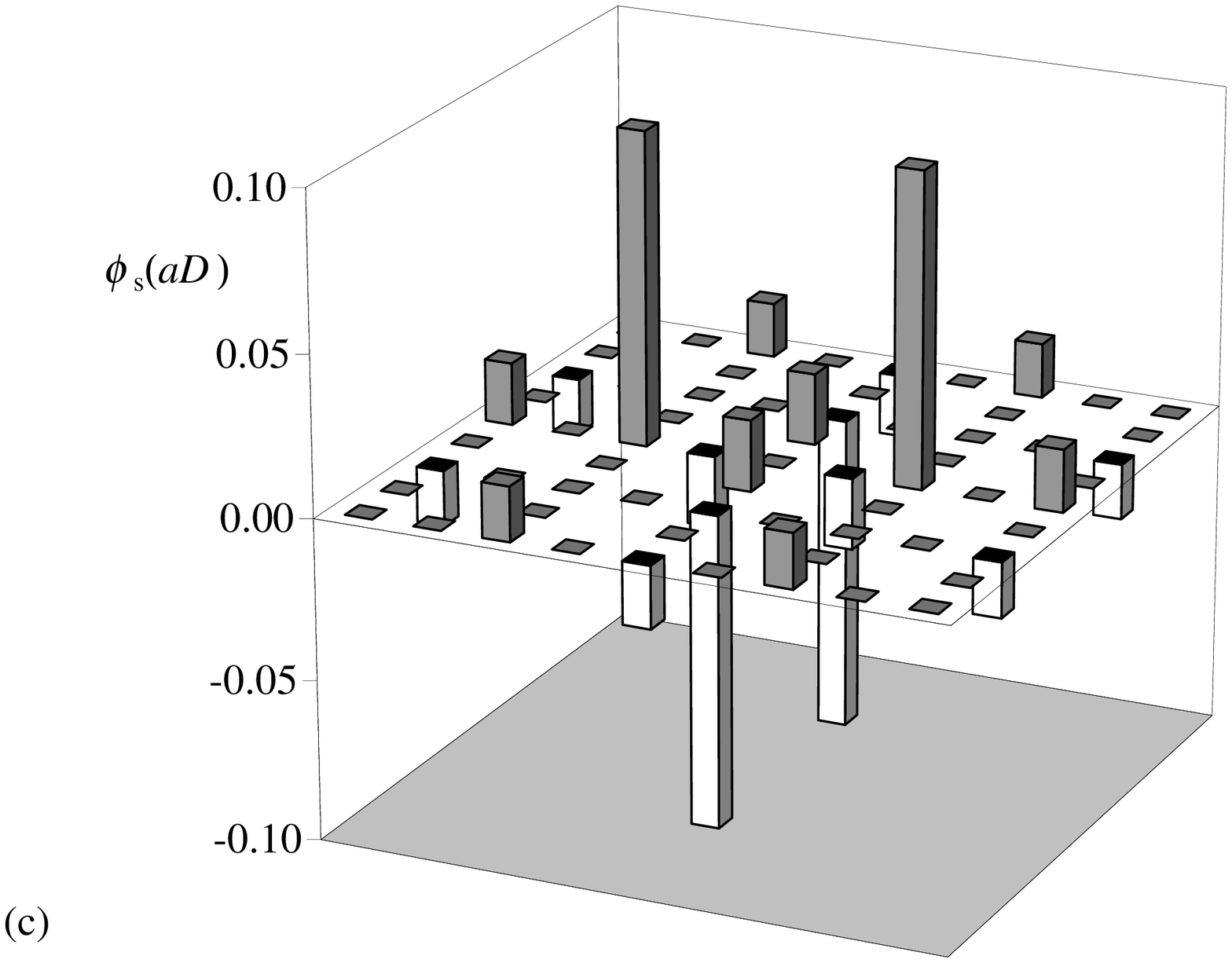}
\includegraphics{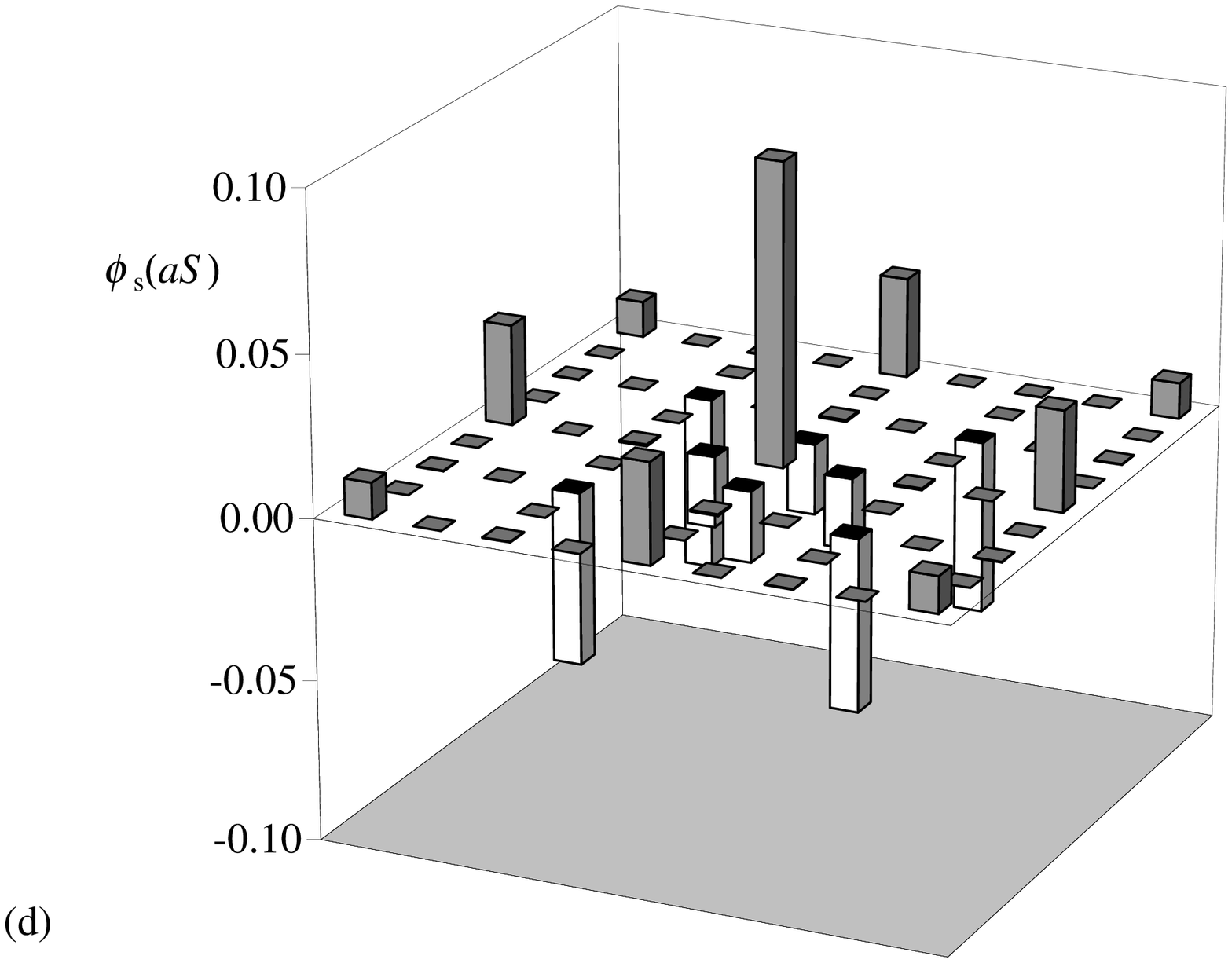}
\end{picture}
\centering
Figure~\ref{tf21}~(continued)
\end{figure}
It is clear that the
wave function is dominated by the d-orbital components; the p-orbital components,
however, play an important role in determining eigenvalues and transition
temperatures, and their neglect is not justified.

\begin{figure}[hbtp]
\begin{picture}(150,85)
\includegraphics{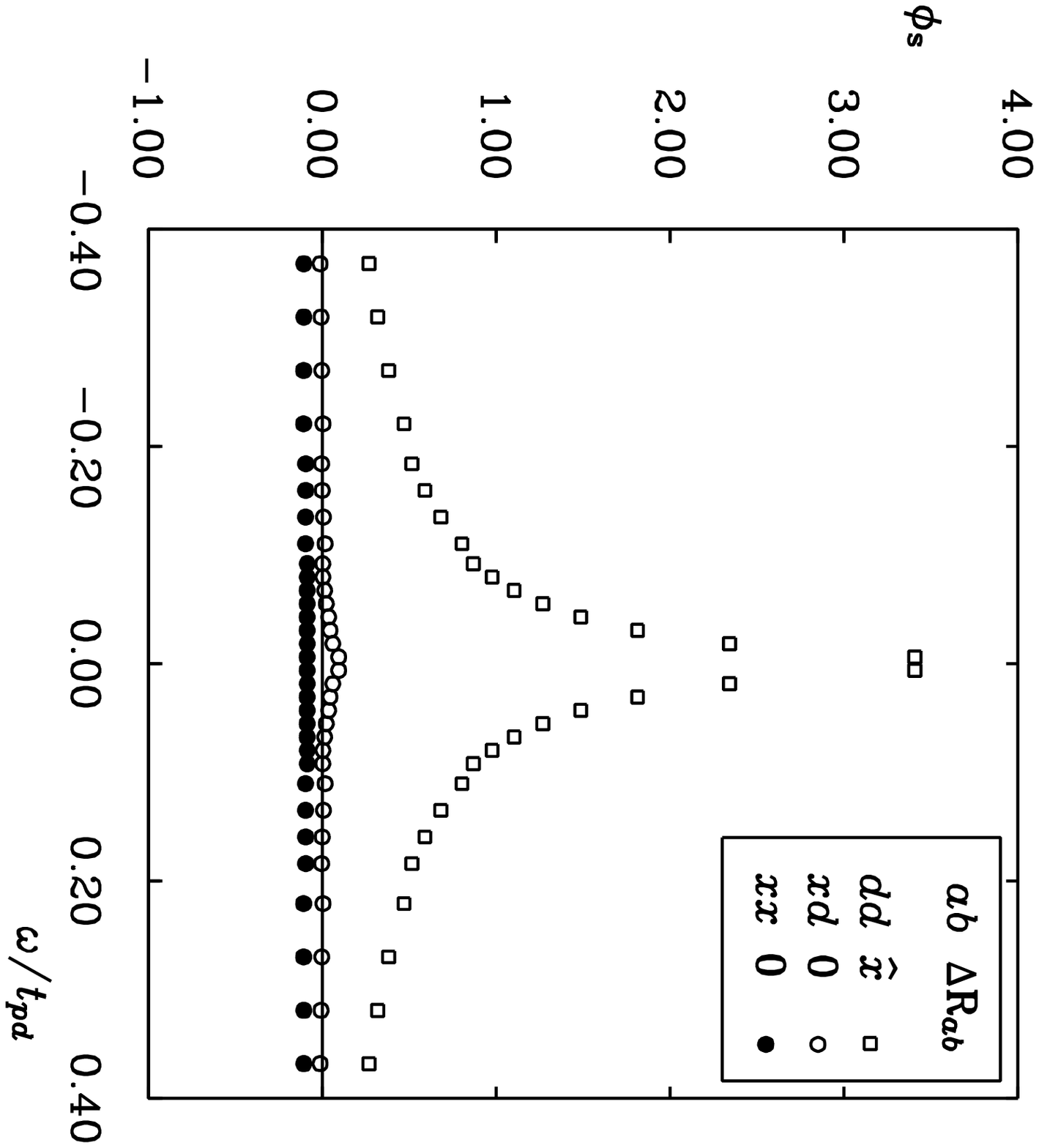}
\end{picture}
\caption{Relative frequency dependence of short-range components of the
$\dxy$ pair wave function for the standard parameter set at $\navg=1.16$. The dominant
component, associated with pairing on near-neighbor Cu 3d$_{\rm x^2-y^2}$ sites,
drops to zero over a range determined by the spin fluctuations.}
\label{tf22}
\end{figure}

Finally the relative frequency dependence of several short-range components of the
pair wave function 
$\phi_{\rm s}({\bf 0};\,ab,\,\Delta{\bf R}_{ab},\,i\omega)$ is displayed in 
Figure~\ref{tf22}. Note that, as in
the one-band Hubbard model,\cite{tr21} the wave function is strongly frequency dependent, falling
rapidly to zero on a scale of $\omega\sim 0.5\tpd$.

\subsection{Eigenvalues and Wave Functions for Particle-Hole Channels}
\label{ch4rese}

A complete FLEX analysis of the particle-hole (i.e., magnetic and charge density)
channels with the same degree of rigor applied in the particle-particle analysis 
is not attempted here. The reasons are as follows: (i)~The version of FLEX considered
in the present work and in EB is based on particle-hole 
exchange. Consequently the particle-particle
vertex functions analyzed in the preceding section contain only 
single-fluctuation-exchange ladders, yet are fully conserving. On the other
hand, a fully conserving calculation of the particle-hole vertex functions within this
approximation scheme requires 
the inclusion of not only single-exchange ladders, but also a class of
double-exchange Aslamazov-Larkin diagrams.\cite{tr18,tr19} Such diagrams constitute the beginning 
of a parquet-like renormalization of the single-exchange ladders.\cite{tr14}
Since this
renormalization is incomplete (and does not improve the consistency of the
particle-hole vertices which appear at different points in the calculation), the
treatment of the Aslamazov-Larkin diagrams is problematic. 
(ii)~In order to treat
the particle-hole vertices on the same footing as the particle-particle, 
the FLEX approximation should include particle-particle exchange diagrams from
the outset (see, e.g., References~\onlinecite{tr14,tr18,tr19}). Such a treatment, while in principle quite
straightforward, exceeds the scope of the present work.

The limitation imposed by these points makes a satisfactory analysis
of the nearly singular magnetic channel impossible in the present work. This is because
both the double-exchange Aslamazov-Larkin 
diagrams and the single-exchange diagrams from the crossed particle-particle channel
are {\sl repulsive} in the magnetic channel. The omission of these contributions
in a naive calculation leads to a drastic 
overestimate of the magnetic eigenvalue (i.e., values larger than unity). A similar
situation has been noted in previous 
FLEX studies of the one-band Hubbard model;\cite{tr4,tr19} in that case the
magnetic eigenvalue drops well below unity when the omitted contributions are 
reinstated. 

Note that these limitations in the treatment of the particle-hole channel do not 
compromise the conserving nature of the FLEX calculation for the particle-particle channel.
(This is not to say that satisfying conservation laws guarantees accuracy:
the overall lack of self-consistency in two-particle vertices which appear at
different points in the FLEX calculation is a broader global concern,\cite{tr14}
which can be remedied only by a more sophisticated parquet-like
analysis. See Section~\ref{ch4sum} for further comments on this point.)

\begin{figure}[hbtp]
\begin{picture}(150,35)
\includegraphics{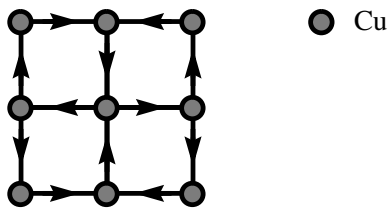}
\end{picture}
\caption{Representation of the circulating charge density currents in an ordered
orbital antiferromagnetic state. The O sites are omitted for clarity.}
\label{tf23}
\end{figure}

Despite these caveats, we feel it important to emphasize in this section a feature of the
physics in the {\sl charge density} channel which has received little attention in
recent years.\cite{tr28} The $\bfQ=0$ $\dxy$ state in the
singlet channel has an analog at $\bfQ\sim (\pi,\,\pi)$ in the density channel. The
presence of such an analog or partner state is familiar in a simpler context: in
the one-band negative-$U$ Hubbard model,\cite{tr29,tr30} the physics near half filling is
dominated by a $\bfQ=0$ s-wave singlet state and a $\bfQ=(\pi,\,\pi)$ charge density state.
At half-filling these states become exactly degenerate, constituting the
components of a Heisenberg-like order parameter. The instability in both the singlet and
density channels is driven by the attractive unrenormalized vertex $-U$. In the
positive-$U$ Hubbard model and the CuO$_2$ model, an analogous pair of potential
instabilities is driven by the exchange of spin fluctuations. In this case the
singlet state of interest has $\dxy$ symmetry. The partner state, which becomes exactly
degenerate with the $\dxy$ singlet at half-filling in the positive-$U$ Hubbard model,
is a $\bfQ=(\pi,\,\pi)$ charge density state which shares the discrete $\dxy$ 
rotational symmetry. This state has been previously considered in both weak- and
strong-coupling studies.\cite{tr24,tr25} We follow Schulz\cite{tr24} in denoting this state 
an ``orbital antiferromagnet''; the name is natural since the state describes 
microscopic currents
which flow around elementary plaquettes in the square lattice, with the direction of
current flow staggered between adjacent plaquettes (see Figure~\ref{tf23}). In strong coupling
the corresponding state\cite{tr25,tr28} has been denoted a ``flux phase.'' 

Away from half-filling in the Hubbard model and at arbitrary fillings in the CuO$_2$
model, the exact symmetry between the $\dxy$ singlet and the orbital antiferromagnet
is broken. Furthermore, with the loss of perfect nesting in the band structure,
the wave vector $\bfQ$ for the optimal charge density state becomes incommensurate.
For example, for the standard parameter set at $\navg=1.16$ the optimal $\bfQ$ vector is
approximately $(1,\,0.875)\pi$.

We have studied the temperature variation of the orbital antiferromagnetic eigenvalue
within an inherently limited approximation: namely, we have included in the charge density
vertex only single-exchange diagrams describing magnetic and density fluctuations.
As noted above in the comments on the magnetic vertex, this approximation is not
entirely satisfactory, since both double-exchange Aslamazov-Larkin diagrams and
single-exchange particle-particle ladders are omitted; however, this approximation does
preserve the crucial feature which determines both the $\dxy$ singlet and orbital
antiferromagnetic eigenvalues, i.e., the exchange of nearly antiferromagnetic spin
fluctuations.

\begin{figure}[hbtp]
\begin{picture}(150,165)
\includegraphics{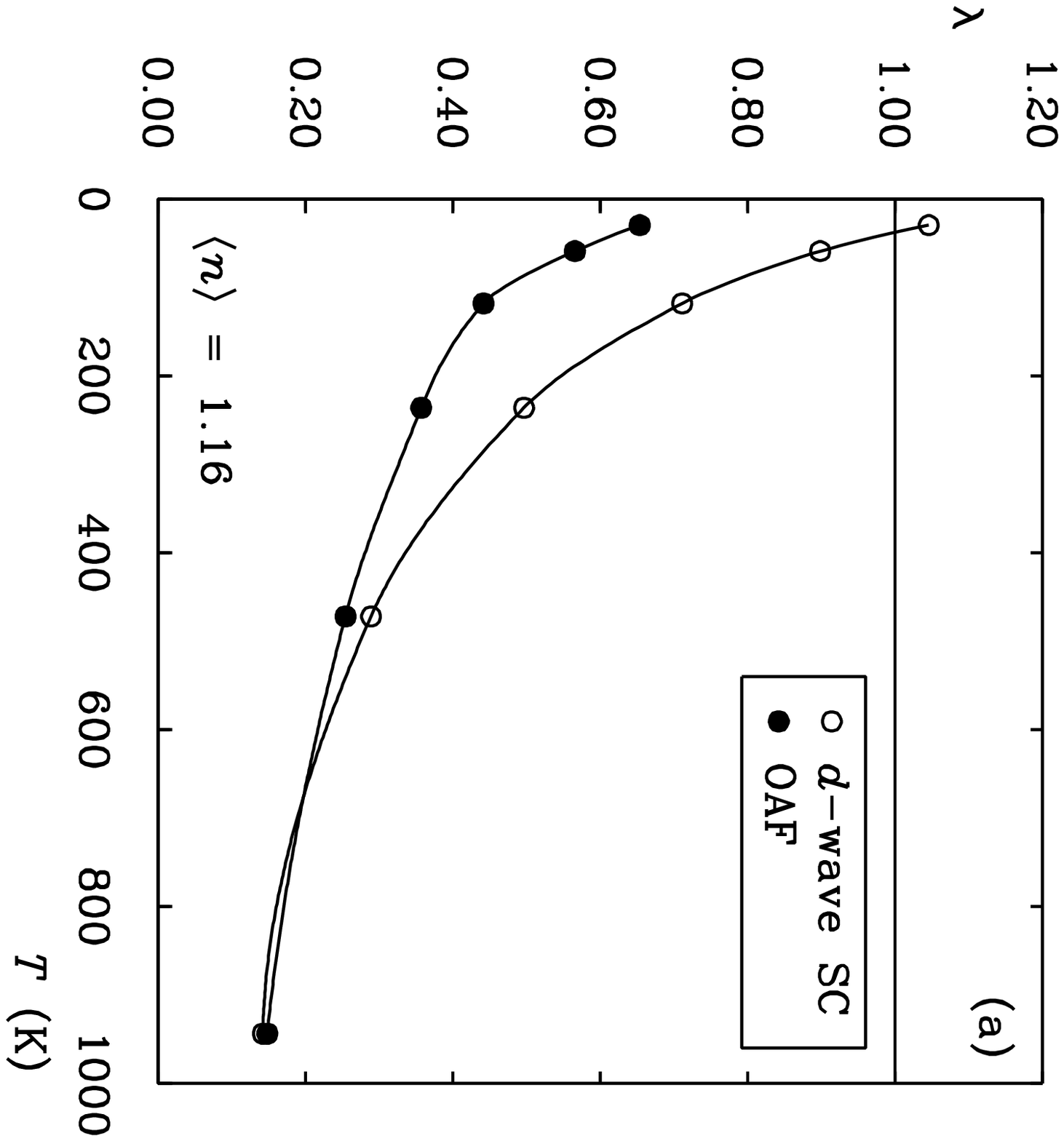}
\includegraphics{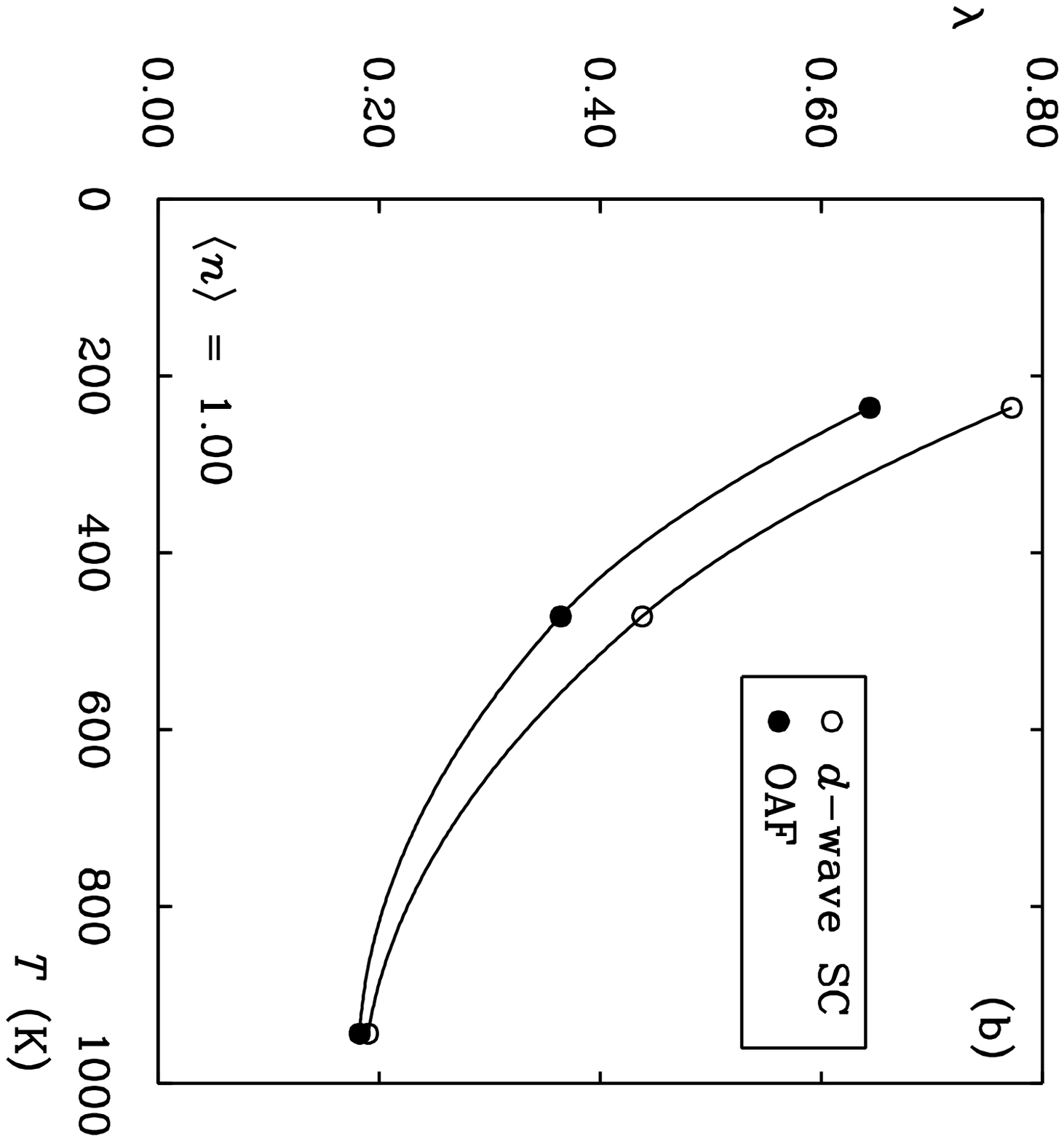}
\end{picture}
\caption{Temperature dependence of the $\dxy$ singlet and 
orbital antiferromagnetic (OAF) eigenvalues
for the standard parameter set. (a)~Results for filling $\navg=1.16$. The optimal wave vector
for the OAF state is in this case $\bfQ=(1,\,0.875)\pi$.
(b)~Results for filling $\navg=1.00$. The optimal wave vector is in this case $\bfQ=(\pi,\,\pi)$.}
\label{tf24}
\end{figure}

The temperature variation of the $\dxy$ singlet and optimal orbital 
antiferromagnetic eigenvalues for the standard parameter set is shown in Figure~\ref{tf24}.
While $\lambda_{\rm\scriptscriptstyle OAF}$ is smaller than $\lambda_{\rm d}$,
the eigenvalues remain very close down to the $\dxy$ transition.
From this analysis it
becomes clear that a fully satisfactory treatment of the model, particularly in the
underdoped regime, must describe the competition between the magnetic, $\dxy$ singlet,
and orbital antiferromagnetic channels. While we have no evidence that the orbital
antiferromagnet is ever actually the dominant instability, it is tempting to speculate
on its relevance, at least in conjunction with the $\dxy$ singlet, for a description of
the anisotropic pseudogap observed in many experimental studies.\cite{tr23}

\begin{figure}[btp]
\begin{picture}(150,110)
\includegraphics{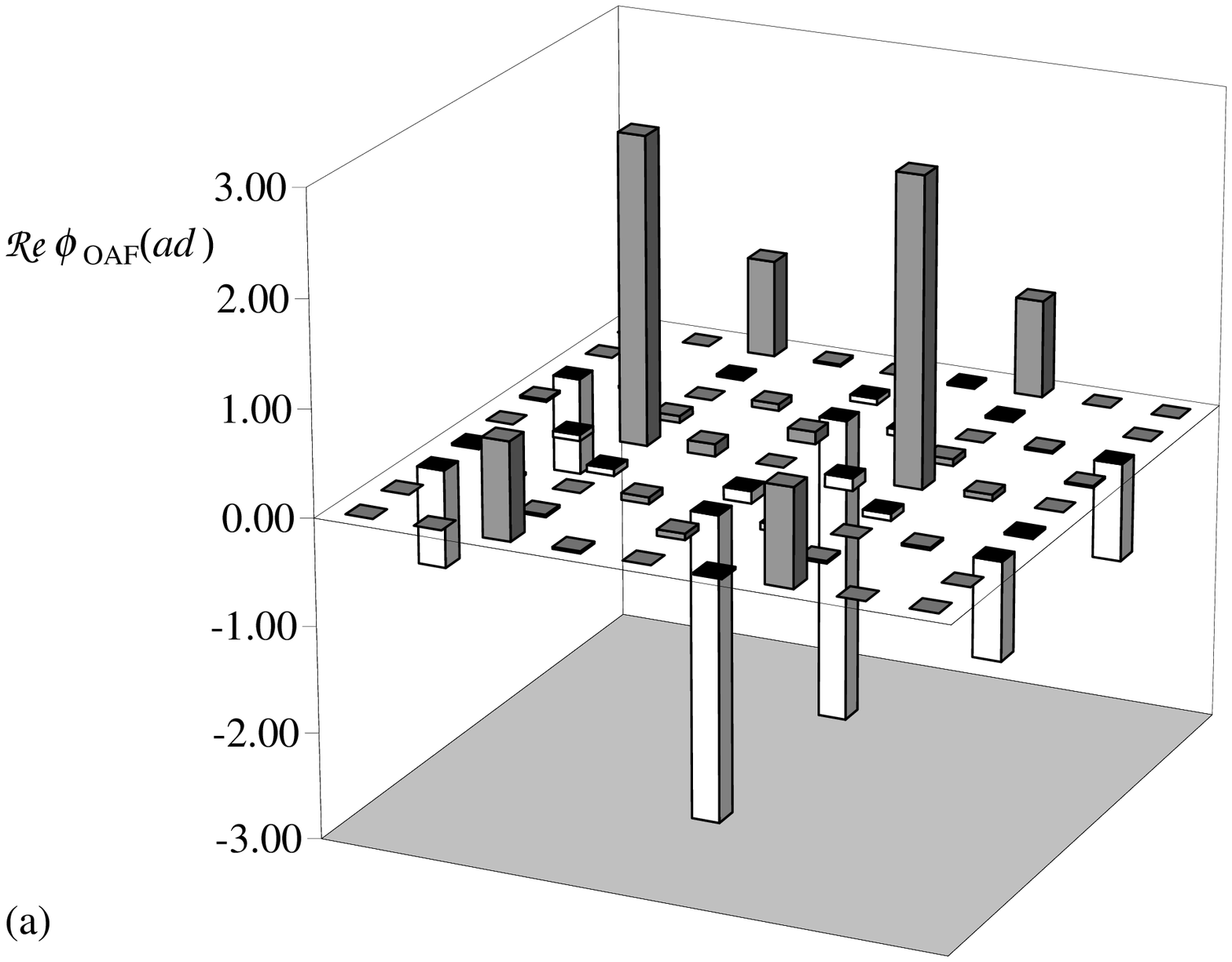}
\end{picture}
\caption{Relative displacement dependence of the dominant minimum-frequency 
($\omega=\pi T$) component of the $\bfQ=(\pi,\,\pi)$ orbital antiferromagnetic wave function,
$\phi_{\rm\scriptscriptstyle OAF}(Q;\,ad,$ $\DeltaR,\,i\pi T)$
for the standard parameter set at $\navg=1.16$. The temperature is $T=\tpd/512$.
As in Figure~\protect\ref{tf21}, the histograms are centered
on a Cu site in the cell with $\DeltaR=0$. In this case the wave function is complex-valued,
even though the eigenvalue is real. Note that $a$ and $\DeltaR$ vary with $b=d$ held
fixed. Note also the fact that $\dxy$ rotational symmetry is manifest for $a=d$ and
arbitrary $\DeltaR$, but that the rotational symmetry is hidden for $a=x$ and $y$, as
discussed at length in the text. (a)~Real part of the wave function.
(b)~Imaginary part of the wave function.}
\label{tf25}
\end{figure}

In Figure~\ref{tf25} we show the spatial variation of the real and imaginary parts of the
$\bfQ=(\pi,\,\pi)$ orbital antiferromagnetic wave function 
$\phi_{\rm\scriptscriptstyle OAF}(Q;\,ad,\,
\Delta\bfR_{ad},\,i\omega=i\pi T)$ for the standard parameter set at $\langle n\rangle=
1.16$ and $T=\tpd/512$.
\begin{figure}[hbtp]
\begin{picture}(150,110)
\includegraphics{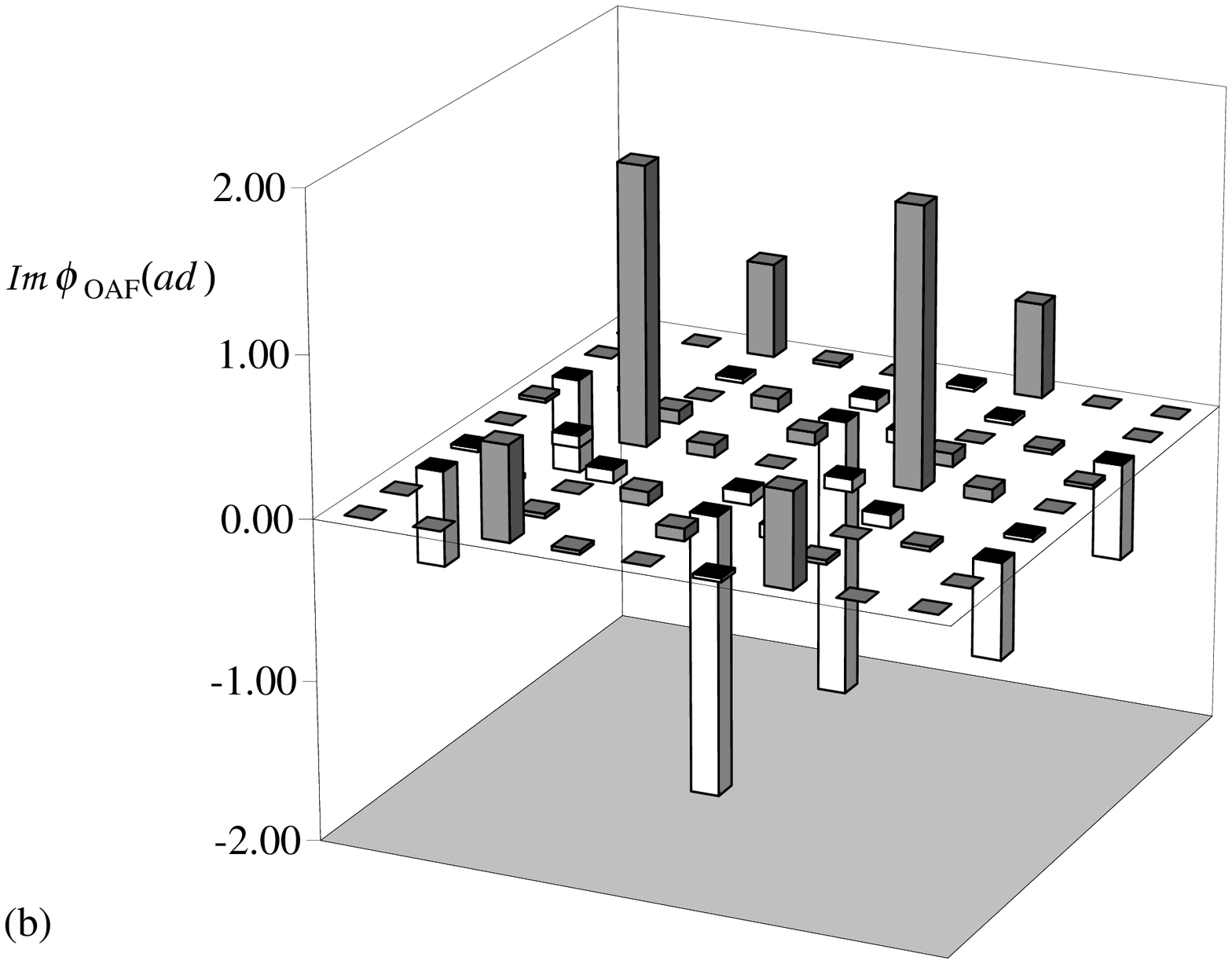}
\end{picture}
\centering
Figure~\ref{tf25}~(continued)
\end{figure}
The $\dxy$ rotational symmetry of the wave function is manifest for
the components with $a=d$, but is hidden for the components with $a=x$ and $y$. This is
true for the following reason: The total wave function takes the form
\begin{eqnarray}
\label{te54}
\lefteqn{\bsp\bsp\sum_{\bfR;\,ab,\,\Delta\bfR_{ab},\,\omega}\!\!\!\!\!\!\!\! 
  e^{i\bfQ\cdot(\bfR+\Delta
  \bfR_{ab})}\,\phi_{\rm\scriptscriptstyle OAF}(Q;\,ab,\,\Delta\bfR_{ab},\,i\omega)}\nn
  &&\times\sum_{\sigma\sigma'}\,\chi_{\rm d}^{\sigma\sigma'}\,\cbar_{a\sigma}
  (\bfR+\Delta\bfR_{ab},\,i(\omega+\Omega))c_{b\sigma'}(\bfR,\, i\omega)\ , 
\end{eqnarray}
with
\begin{eqnarray}
\label{te55}
  \chi_{\rm d}&=&{1\over\sqrt 2}\left( \matrix{ 1 & 0 \cr 0 & 1 \cr}\right)\nn
   \bfQ&=&(\pi,\,\pi) \nn
   \Omega&=&0\ .  
\end{eqnarray}
It is convenient to adopt the shorthand $\widetilde a$ for the rotated image of orbital
$a$ and $\widetilde\bfR_{\widetilde a}$ for the rotated image of $\bfR_a$, the unit-cell
location of orbital $a$.
When the wave function coordinates (unit-cell and orbital labels) are rotated, it is
guaranteed that the compound d-orbital label $(a,\,\bfR_a)=(d,\,\bfR)$ maps to
$(\widetilde a,\,\widetilde\bfR_{\widetilde a})=(d,\,\widetilde\bfR)$, where $\widetilde\bfR$
is the rotated image of $\bfR$. However, under successive rotations, the p$_{\rm x}$-orbital
label $(x,\,\bfR)$ maps to $(y,\,\widetilde\bfR)$ (rotation through $\pi/2$);
$(x,\,\widetilde\bfR-\widehat x)$ (rotation through $\pi$); and 
$(y,\,\widetilde\bfR-\widehat y)$ (rotation through $3\pi/2$). A similar set of 
transformations holds for the p$_{\rm y}$-orbital label. Discrete $\dxy$ symmetry requires that
\begin{eqnarray}
\label{te56}
\lefteqn{ \bsp\bsp e^{i\bfQ\cdot\widetilde\bfR_{\widetilde a}}\,\phi_{\rm\scriptscriptstyle OAF}
  (Q;\,\widetilde a\widetilde b,\,\widetilde\bfR_{\widetilde a}-\widetilde\bfR_{\widetilde b},
  \,i\omega)\ =} \nn
  && e^{2i\theta}\,e^{i\bfQ\cdot\bfR_a}\,\phi_{\rm\scriptscriptstyle OAF}(Q;\,
  ab,\,\bfR_a-\bfR_b,\,i\omega)\ ,  
\end{eqnarray}
where
\begin{eqnarray}
\label{te57}
 \bfR_b&=&\bfR \nn
            \bfR_a&=&\bfR+\Delta\bfR_{ab} \nn
            \theta&=&0,\ \pi/2,\ \pi,\ 3\pi/2\ .  
\end{eqnarray}
When the phase factors $e^{i\bfQ\cdot\widetilde\bfR_{\widetilde a}}$ and $e^{i\bfQ\cdot\bfR_a}$
are equal, the $\dxy$ symmetry is manifest in $\phi_{\rm\scriptscriptstyle OAF}$, i.e.,
\begin{equation}
\label{te58}
\phi_{\rm\scriptscriptstyle OAF}(Q;\,\widetilde a\widetilde b,\,\widetilde\bfR_{\widetilde
a}-\widetilde\bfR_{\widetilde b},\,i\omega)\ =\ e^{2i\theta}\,\phi_{\rm\scriptscriptstyle OAF}
(Q;\,ab,\,\bfR_a-\bfR_b,\,i\omega)\ . 
\end{equation}
However, when 
\begin{equation}
\label{te59a}
e^{i\bfQ\cdot\widetilde\bfR_{\widetilde a}}\ =\ -e^{i\bfQ\cdot\bfR_a}\ , 
\end{equation}
the $\dxy$ symmetry is hidden, i.e.,
\begin{equation}
\label{te59b}
\phi_{\rm\scriptscriptstyle OAF}(Q;\,\widetilde a\widetilde b,\,\widetilde\bfR_{\widetilde
a}-\widetilde\bfR_{\widetilde b},\,i\omega)\ =\ -e^{2i\theta}\,\phi_{\rm\scriptscriptstyle OAF}
(Q;\,ab,\,\bfR_a-\bfR_b,\,i\omega)\ . 
\end{equation}
This accounts for the seemingly anomalous transformation properties of the $\phi_{xd}$ and
$\phi_{yd}$ components of the wave function in Figure~\ref{tf25}.

\begin{figure}[hbtp]
\begin{picture}(150,165)
\includegraphics{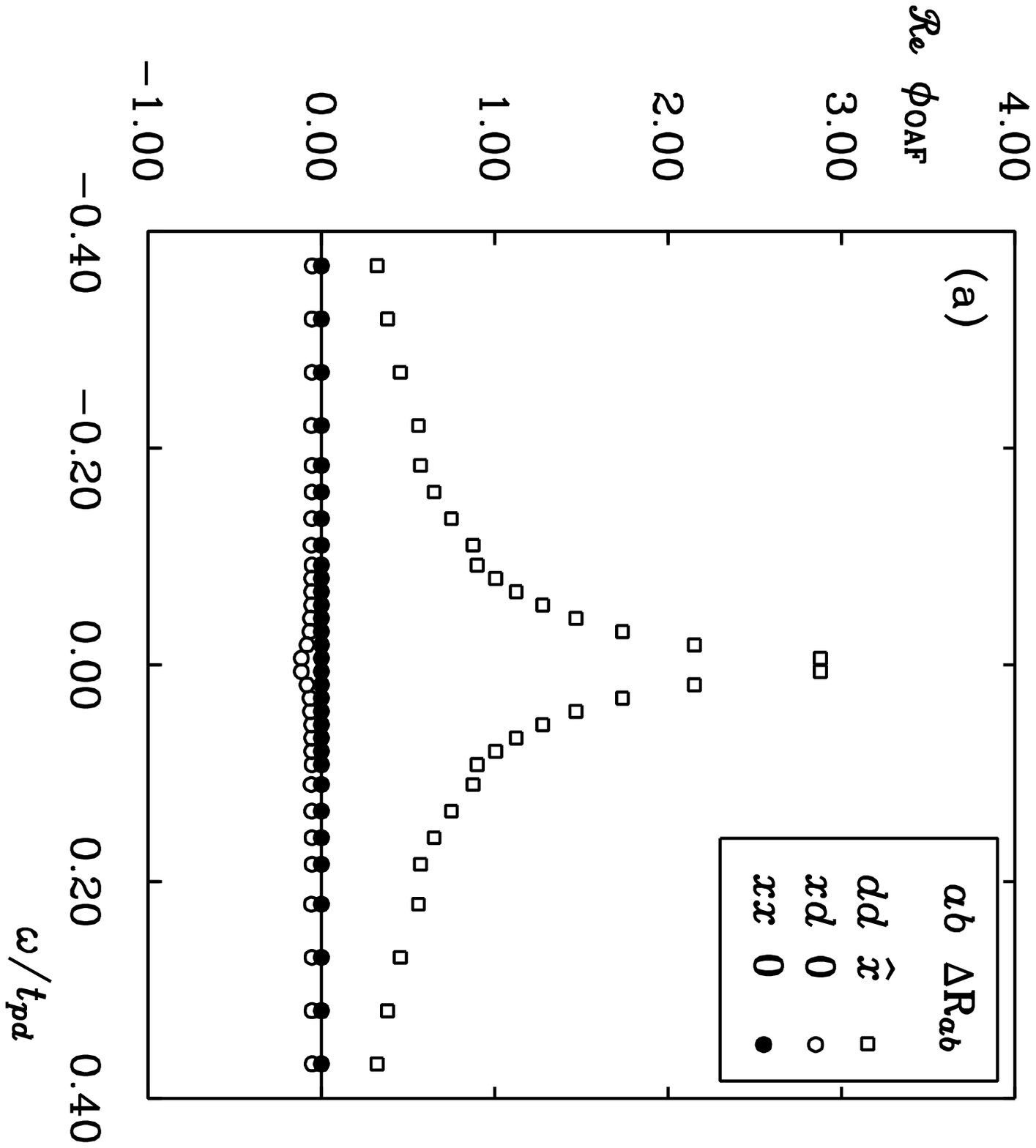}
\includegraphics{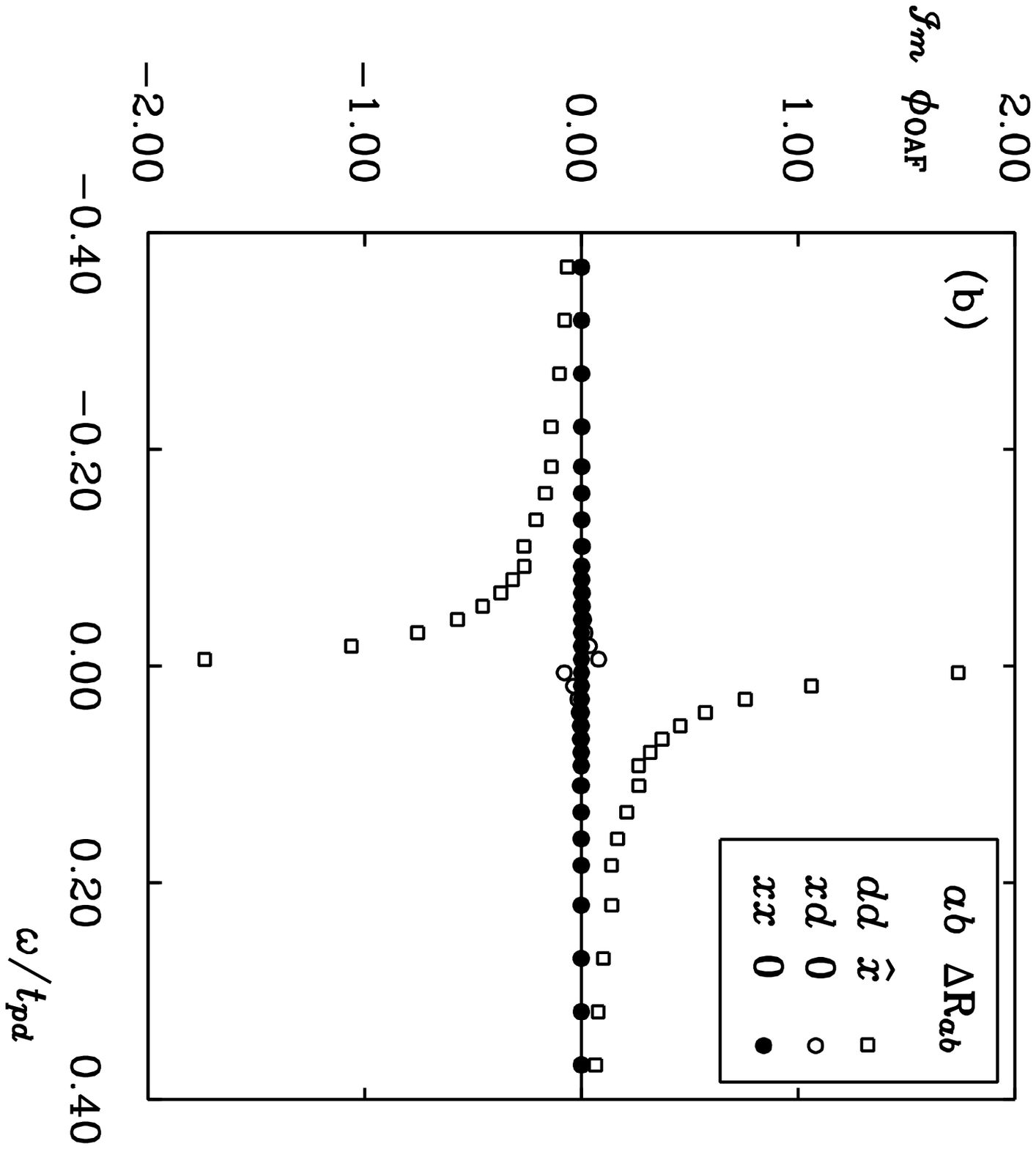}
\end{picture}
\caption{Relative frequency dependence of short-range components of the
$\bfQ=(\pi,\,\pi)$ 
orbital antiferromagnetic wave function for the standard parameter set at $\navg=1.16$. 
As in Figure~\protect\ref{tf22}, the component associated with near-neighbor Cu 3d$_{\rm x^2-y^2}$ sites
drops to zero over a range determined by the spin fluctuations. (a)~Real part of the
wave function. (b)~Imaginary part of the wave function.}
\label{tf26}
\end{figure}

The frequency dependence of several short-range components of the $\bfQ=(\pi,\,\pi)$
orbital antiferromagnetic 
wave function is illustrated in Figure~\ref{tf26}. Note that the wave function is in this case
intrinsically complex, although the eigenvalue is real.
While the Pauli Principle does not
dictate the transformation properties under $\omega\rightarrow -\omega$ in this case,
the overall phase of the wave function for $\bfQ=(\pi,\,\pi)$ and $\Omega=0$
may still be chosen such that
\begin{equation}
\label{te60}
 \phi_{\rm\scriptscriptstyle OAF}(Q;\,ab,\,\DeltaR,\,-i\omega)^*
  \ =\ \phi_{\rm\scriptscriptstyle OAF}(Q;\,ab,\,\DeltaR,\,i\omega)\ . 
\end{equation}
This follows from a basic symmetry of the eigenvalue problem for $K^{\rm ph}_{\rm r}\equiv
\Gamma^{\rm ph}_{\rm r}G^{\rm ph}$ in Equation~(\ref{te32}), viz.,
\begin{eqnarray}
\label{te61}
\lefteqn{\bsp\bsp K^{\rm ph}_{\rm r}(Q;\,m_1m_2,\,\DeltaR_{12},\,i\omega;\,
m_3m_4,\,\DeltaR_{34},\,i\omega')^*\ =}\nn
&&              K^{\rm ph}_{\rm r}(-Q;\,m_1m_2,\,\DeltaR_{12},\,-i\omega;\,
m_3m_4,\,\DeltaR_{34},\, -i\omega')\ .  
\end{eqnarray}
This implies that if $\phi(Q;\,ab,\,\DeltaR,\,i\omega)$ is an eigenfunction 
with eigenvalue $\lambda(Q)$ for total momentum-frequency $Q$, 
then $\phi(Q;\,ab,\,\DeltaR,\,-i\omega)^*$ is an eigenfunction with eigenvalue $\lambda(Q)^*$ 
for total momentum-frequency $-Q$. For the case of interest here $Q=-Q$, 
the eigenfunction $\phi(Q;\,ab,\,\DeltaR,\,i\omega)$ is non-degenerate, and the 
eigenvalue is real. Thus,
\begin{equation}
\label{te62}
\phi(Q;\,ab,\,\DeltaR,\,-i\omega)^*\ =\ \alpha\,\phi(Q;\,ab,\,\DeltaR,\,i\omega)
  \ , 
\end{equation}
with $\alpha$ a complex constant. The constant may 
always be set equal to unity by a simple rescaling of $\phi$, leading to the 
symmetry relation in Equation~(\ref{te60}). Symmetries for general 
$\bf Q$ near $(\pi,\,\pi)$ may be examined by an extension of this argument.
Note finally the overall similarity of the $\dxy$ singlet wave function plotted in Figures~\ref{tf21}--\ref{tf22}
and the
orbital antiferromagnetic wave function plotted in Figures~\ref{tf25}--\ref{tf26}, with respect to both
spatial and frequency dependence.

%% file: sum.tex
Our results demonstrate that the fluctuation exchange approximation provides
reasonable results for both the magnitude and doping dependence of the
$\dxy$ transition temperature in the overdoped regime, $\navg-1>0.16$. 
While the level of quantitative agreement between FLEX and experiment\cite{tr26}
is almost certainly fortuitous, it is important to emphasize several points in
this regard: (i)~For a wide range of model parameters clustered around the
standard LDA set,\cite{tr9} FLEX transition temperatures are predicted in the range of 10 to
100 K. It is by no means obvious that this should be so, i.e., one might have 
imagined obtaining a range of transitions stretching over several orders of
magnitude. (ii)~The continued rise of the FLEX $\dxy$ eigenvalue 
for values of $\navg$ approaching unity is consistent with previous Monte Carlo 
studies,\cite{tr19}
which have demonstrated enhanced $\dxy$ correlations even in regions where
long-range magnetic order is being established.  
It appears clear that a more sophisticated
approach\cite{tr14} is essential to resolve the competition between the incipient
instabilities in the antiferromagnetic spin, $\dxy$ singlet, {\sl and}
orbital antiferromagnetic channels in the underdoped regime $\navg\rightarrow 1.0$. 

As emphasized in Section~\ref{ch4res}, the presence of large eigenvalues in the
orbital antiferromagnetic channel is an unambiguous result of our analysis, despite the
technical limitations of our approach for the particle-hole channels. It is important to
note that the orbital antiferromagnetic channel becomes degenerate with the $\dxy$ singlet 
to form a Heisenberg-like order
parameter in models with exact particle-hole symmetry (such as the half-filled one-band
Hubbard model). While the breaking of particle-hole
symmetry in the standard CuO$_2$ model apparently favors the $\dxy$ singlet at 
half-filling, it seems clear that the orbital
antiferromagnet must be retained in any analysis which aims at a quantitative
description of the region near $\navg=1$.

Finally, as a more general comment, the present study demonstrates the feasibility of
extending the FLEX instability analysis to models with an increasing degree of realism.
While the principal shortcoming of FLEX, viz., the lack of self-consistency at the
two-body level, remains a separate concern, it is also clear that progress toward a
truly predictive many-body theory demands the ability to incorporate realistic details of
lattice and interaction structure. A natural next step in this direction is the
analysis of a one-band model with longer-range interactions. The general formalism
developed in the present work and in EB (in particular, the use of a
real-space basis set for relative coordinates) provides a calculationally feasible
framework for such a study.

%% file: ack.tex
This work was supported in part by the National Science Foundation under grants
DMR95--20636 and PHY94--07194; and by the Department of Energy under grant
85--ER--45197. NEB gratefully acknowledges the hospitality of the Institute
for Theoretical Physics where this work was begun. We also thank D.~J.
Scalapino, N. Bulut, and C.-H. Pao for a series of fruitful discussions.

%% file: preprint.bbl
\begin{thebibliography}{99}
\bibliographystyle{prsty}

\bibitem{tr1}
See, e.g., the extensive review by D.~J. Scalapino, Physics Reports
{\bf 250}, 329 (1995).

\bibitem{tr2}
D.~J. Scalapino, E. Loh, and J.~E. Hirsch, Phys. Rev. B {\bf 34},
8190 (1986); K. Miyake, S. Schmitt-Rink, and C.~M. Varma, Phys. Rev. B {\bf 34}, 6554 (1986).

\bibitem{tr3}
N.~E. Bickers, D.~J. Scalapino, and R.~T. Scalettar, Int. J. Mod. Phys.
B {\bf 1}, 687 (1987).

\bibitem{tr4}
N.~E. Bickers, D.~J. Scalapino, and S.~R. White, Phys. Rev. Lett. {\bf 62},
961 (1989).

\bibitem{tr5}
A.~E. Ruckenstein, P.~J. Hirschfeld, and J. Appel, Phys. Rev. B {\bf 36},
857 (1987); G. Kotliar, Phys. Rev. B {\bf 37}, 3664 (1988).

\bibitem{tr6}
See also the phenomenological studies by P. Monthoux, A.~V. Balatsky, and
D. Pines, Phys. Rev. Lett. {\bf 67}, 3448 (1991); P. Monthoux and D. Pines,
{\em ibid.\/} {\bf 69}, 961 (1992); S. Lenck, J.~P. Carbotte, and R.~C. Dynes, Phys. Rev. B {\bf 49}, 4176 (1994).

\bibitem{tr7}
V.~J. Emery, Phys. Rev. Lett. {\bf 58}, 2794 (1987).

\bibitem{tr8}
F.~C. Zhang and T.~M. Rice, Phys. Rev. B {\bf 37}, 3759 (1988).

\bibitem{tr9}
M.~S. Hybertsen, M. Schl$\rm\ddot u$ter, and N.~E. Christensen,
Phys. Rev. B {\bf 39}, 9028 (1989).

\bibitem{tr10}
A.~K. McMahan, R.~M. Martin, and S. Satpathy, Phys. Rev. B {\bf 38},
6650 (1988).

\bibitem{tr11}
E.~B. Stechel and D.~R. Jennison, Phys. Rev. B {\bf 38}, 4632 (1988).

\bibitem{tr12}
G\"{o}khan Esirgen and N.~E. Bickers, Phys. Rev. B {\bf 55}, 2122 (1997).

\bibitem{tr13}
An earlier normal-state study of the CuO$_2$ model with $\Upp=\Upd=0$
was carried out by J. Luo and N.~E. Bickers, Phys. Rev. B {\bf 47}, 12153 (1993).

\bibitem{tr14}
N.~E. Bickers, in {\em Numerical Methods for Lattice Quantum Many-Body
Problems\/} (Addison Wesley, Reading MA, in press).

\bibitem{tr15}
Such a basis set was first employed in the context of the CuO$_2$ model
by P.~B. Littlewood, C.~M. Varma, S. Schmitt-Rink, and E. Abrahams, Phys. Rev. B {\bf 39},
12371 (1989). See also Y. Bang, K. Quader, E. Abrahams, and P.~B. Littlewood, Phys. Rev. B {\bf 42}, 4865 (1990).

\bibitem{tr16}
D.~C. Sorensen, SIAM J. Matr. Anal. Apps. {\bf 13}, 357 (1992); B.~N. Parlett
and Y. Saad, Linear Algebra and its Applications {\bf 88/89}, 575 (1987).
A publicly available version of the Lanczos-Arnoldi algorithm resides in the
ARPACK Internet library.

\bibitem{tr17}
G. Baym and L.~P. Kadanoff, Phys. Rev. {\bf 124}, 287 (1961); G. Baym,
{\em ibid.\/} {\bf 127}, 1391 (1962).
See also the discussion in Reference~\onlinecite{tr14}.

\bibitem{tr18}
N.~E. Bickers and D.~J. Scalapino, Ann. Phys. (N.~Y.) {\bf 193}, 206 (1989).

\bibitem{tr19}
N.~E. Bickers and S.~R. White, Phys. Rev. B {\bf 43}, 8044 (1991).

\bibitem{tr20}
Chien-Hua Pao and N.~E. Bickers, Phys. Rev. B {\bf 49}, 1586 (1994).

\bibitem{tr21}
Chien-Hua Pao and N.~E. Bickers, Phys. Rev. Lett {\bf 72}, 1870 (1994);
Phys. Rev. B {\bf 51}, 16310 (1995).

\bibitem{tr22}
See, e.g., C. Itzykson and J.-M. Drouffe, {\em Statistical Field Theory\/},
(Cambridge University Press, 1989), Vol. 1.

\bibitem{tr23}
See, e.g., the following  and references therein: Z.-X. Shen and
J.~R. Schrieffer, Phys. Rev. Lett. {\bf 78}, 1771 (1997) (ARPES); R. Nemetschek \etal,
Phys. Rev. Lett. {\bf 78}, 4837 (1997) (Raman); and G.~V.~M. Williams \etal, Phys. Rev.
Lett. {\bf 78}, 721 (1997) (NMR).

\bibitem{tr24}
H.~J. Schulz, Phys. Rev. B {\bf 39}, 2940 (1989).

\bibitem{tr25}
I. Affleck and J.~B. Marston, Phys. Rev. B {\bf 37}, 3774 (1988).

\bibitem{tr26}
M.~K. Crawford \etal, Phys. Rev. B {\bf 41}, 282 (1990).

\bibitem{tr27}
See, e.g., D.~J. Scalapino, in {\em Superconductivity\/}, edited by R.~D. Parks
(Marcel Dekker, New York, 1969), Vol. 1.

\bibitem{tr28}
Note, however, the recent re-analysis of the flux phase by
X.-G. Wen and P.~A. Lee, Phys. Rev. Lett. {\bf 76}, 503 (1996).

\bibitem{tr29}
See, e.g., A. Moreo and D.~J. Scalapino, Phys. Rev. Lett. {\bf 66}, 946 (1991),
and references therein.

\bibitem{tr30}
J. Luo and N.~E. Bickers, Phys. Rev. B {\bf 48}, 15983 (1993).

\end{thebibliography}
